\documentclass[submission, Phys]{SciPost}

\usepackage{amsmath,graphicx}
\usepackage[utf8]{inputenc}
\usepackage[T1]{fontenc}
\usepackage{amsmath,amssymb,amsfonts,amsthm,mathtools}
\usepackage{rotating}

\hypersetup{
    colorlinks,
    linkcolor={red!50!black},
    citecolor={blue!50!black},
    urlcolor={blue!80!black}
}

\usepackage{tikz}
\usepackage{tikz-feynhand}
\usetikzlibrary{shapes.misc}
\usetikzlibrary{decorations.pathmorphing}

\usepackage{tikz}

\newcommand{\nc}{\newcommand}
\nc{\braoprket}[3]{\langle#1|#2|#3\rangle}
\nc{\opn}[1]{\operatorname{#1}}
\nc{\avg}[1]{\langle#1\rangle}
\nc{\ketbrasame}[1]{|#1\rangle\!\langle#1|}
\nc{\ket}[1]{|#1\rangle}
\nc{\bra}[1]{\langle#1|}
\nc{\braket}[1]{\langle|#1 | \rangle }
\nc{\swap}{\opn{SWAP}}
\nc{\E}{\mathbb{E}}
\nc{\Var}{\opn{Var}}
\nc{\dg}{\dagger}

\newcommand{\panel}[3]{%
  \begin{tikzpicture}[inner sep=0pt, outer sep=0pt]
    \node[anchor=south west, inner sep=1.5pt] (img) at (0,0)
      {\includegraphics[width=\dimexpr #2-3pt\relax]{#1}}; 
    \node[anchor=north west, font=\bfseries, fill=white, rounded corners=0pt, inner sep=0pt]
      at ([xshift=0pt,yshift=-0.1pt]img.north west) {#3};
  \end{tikzpicture}%
}

\newcommand{\panelH}[3]{
	\begin{tikzpicture}[inner sep=0pt, outer sep=0pt,
	baseline=(current bounding box.north)]
	\node[anchor=south west, inner sep=1.5pt] (img) at (0,0)
	{\includegraphics[height=\dimexpr #2-3pt\relax,keepaspectratio]{#1}};
	\path ([yshift=0.6ex]img.north west) -- ([yshift=0.6ex]img.north east);
	\node[anchor=north west, font=\bfseries, fill=white,
	inner sep=0.6pt, text height=1.3ex, text depth=0.25ex]
	at ([yshift=-0.4ex]img.north west) {#3};
	\end{tikzpicture}%
}

\usepackage[normalem]{ulem}

\begin{document}

\begin{center}{\Large \textbf{
$\mathcal{PT}$ symmetry-enriched non-unitary criticality
}}\end{center}

\begin{center}

Kuang-Hung Chou\textsuperscript{1},
Xue-Jia Yu\textsuperscript{2,3,4*},
Po-Yao Chang\textsuperscript{1,5$\dagger$}
\end{center}

\begin{center}
{\bf 1} Department of Physics, National Tsing Hua University, Hsinchu 30013, Taiwan
\\
{\bf 2} Eastern Institute of Technology, Ningbo 315200, China
\\
{\bf 3} Department of Physics, Fuzhou University, Fuzhou 350116, Fujian, China
\\
{\bf 4} Fujian Key Laboratory of Quantum Information and Quantum Optics,
College of Physics and Information Engineering,
Fuzhou University, Fuzhou, Fujian 350108, China
\\
{\bf 5} Yukawa Institute for Theoretical Physics, Kyoto University, Kyoto 606-8502, Japan
\\
* xuejiayu@eitech.edu.cn, $\dagger$ pychang@phys.nthu.edu.tw
\end{center}

\begin{center}
\today
\end{center}


\section*{Abstract}
{\bf
The interplay between topology and quantum criticality has given rise to the notion of symmetry-enriched criticality, which has attracted considerable attention in recent years. In this Letter, we demonstrate that parity–time ($\mathcal{PT}$) symmetry enriches non-Hermitian critical points, establishing a topologically distinct class of non-unitary criticality. Through the analytic solution of $\mathcal{PT}$-symmetric free-fermion models, we reveal a new family of critical points that are topologically nontrivial and host robust edge modes. Crucially, these points cannot be adiabatically connected to trivial ones without breaking $\mathcal{PT}$-symmetry or crossing a multicritical point, and distinct from Hermitian counterparts. We further show that, at these $\mathcal{PT}$-symmetry-enriched critical points, conformal scaling of the entanglement entropy necessarily comes with a quantized imaginary subleading term, whose quantization is set by the number of boundary modes in the reduced density matrix. This term is robust against $\mathcal{PT}$-symmetric disorder and interactions, and admits an interpretation as the Affleck–Ludwig $g$-factor associated with the boundary states.
These phenomena are shown to arise from a generalized mass inversion unique to non-Hermitian criticality.
}

\vspace{10pt}
\noindent\rule{\textwidth}{1pt}
\tableofcontents\thispagestyle{fancy}
\noindent\rule{\textwidth}{1pt}
\vspace{10pt}

\section{Introduction}
\label{sec:intro}
Non-Hermitian quantum systems have recently attracted considerable attention due to their unique properties beyond Hermitian counterparts~\cite{ashida2020non,El-Ganainy2018NP,Kawabata2019PRX,Ashida2017NC,Xiao2019PRL} and are closely related to diverse experimental platforms~\cite{Xiao2020NP,Zenuer2015PRL,Gao2015Nature,Cao2015RMP,Xiao2021PRL}. The characterization of quantum phases in non-Hermitian systems is now well established~\cite{Yokomizo2019PRL,Kawabata2020PRB,Wang2024PRX,Guo2023PRR,Wu2020PRB,Shen2023PRL}, revealing distinctive phenomena such as the non-Hermitian skin effect~\cite{Okuma2020PRL,Li2020NC,Zhang2022NC,Song2019PRL,Zhang2021ObsNC,Li2024NC} and non-Hermitian topological phases~\cite{Gong2018PRX,okuma2023non,Bergholtz2021RMP,Yao2018PRL,Shen2018PRL,Lieu2018PRB,Liu2019PRL,Esaki2011PRB,Yao2018EdgePRL,Leykam2017PRL,Lee2020PRB,Lee2016PRL,Xue2020PRL}. In contrast, phase transitions in non-Hermitian systems—typically described by non-unitary conformal field theories (CFTs)—remain less understood, although they have recently drawn growing interest from the perspective of quantum entanglement~\cite{Chang2020PRR,li2025impurityinducednonunitarycriticality,tu2023general,Hamazaki2019PRL,Chen_2024,Guo_2021,Gopalakrishnan2021PRL,Loic2019SciPost,Fossati2023PRB,Chen2021SciPost,Chen2022PRBL,Rottoli_2024,Lee2022PRL,zou2024experimental,liu2024non,xue2025topologicallyprotectednonhermitiansupervolumelaw,Chang2023SciPostCore,Yang2024SciPost,Shimizu2025PRB,Tu2022SciPost,Zhou2022PRR,yi2023exceptionalentanglementnonhermitianfermionic,lu2025biorthogonalquenchdynamicsentanglement,Zou2025PRB,Bianchini_2015,zhou2024universalnonhermitianflowonedimensional,fan2025simulatingnonunitaryyangleeconformal,Fahri2021Science,Longhi2019PRL,Matsumoto2020PRL,Kawabata2023PRX,Arouca2020PRB,Yu2024PRL_b,Li2024PRB,Guo2022PRA,Yang2024PRL,Li_2025,Li2025PRA,Li2023PRA}.

On a different front, the universality class of quantum phase transitions can be further enriched by global symmetries, giving rise to topologically distinct universality classes, now referred to as symmetry-enriched quantum criticality~\cite{Verresen2021PRX,Verresen2018PRL,Jones2019JSP,verresen2020topologyedgestatessurvive,Duque2021PRB,Yu2022PRL,Ye2022SciPost,Mondal2023PRB,Jones2023PRL,Choi2024PRB,Yu2024PRB,Zhong2024PRA,Zhong2025PRB,Zhou2025CP,Li2025PRB,Wang2022SciPost,Wei2024JHEP,rey2025incommensurategaplessferromagnetismconnecting} or gapless symmetry-protected topological states~\cite{YU20261,Keselman2015PRB,Scaffidi2017PRX,JIANG2018753,Parker2018PRB,Hidaka2022PRB,verresen2024higgscondensatessymmetryprotectedtopological,thorngren2023higgscondensatessymmetryprotectedtopological,Thorngren2021PRB,Umberto2021SciPost,Wen2023PRB,Nathanan2023SciPost,Nathanan2023SciPost_b,Yu2024PRL,Zhang2024PRA,yu2025gaplesssymmetryprotectedtopologicalstates,Su2024PRB,Prembabu2024PRB,ando2024gauginglatticegappedgaplesstopological,bhardwaj2024hassediagramsgaplessspt,Li2025SciPost,Li2024SciPost,Wen2025PRB,bhardwaj2025gaplessphases21dnoninvertible,wen2025topologicalholography21dgapped,Andrea2025SciPost,Flores2025PRL,yang2025deconfinedcriticalityintrinsicallygapless,tan2025exploringnontrivialtopologyquantum,Yang2025CP,Huang2025SciPost,Huang2025PRB,wen2024topologicalholographyfermions,wen2025stringcondensationtopologicalholography}. The discovery of topological physics in quantum critical systems opens new avenues for classifying phase transitions within the same universality class, fundamentally enriching the textbook understanding of phase transitions. Remarkably, the coexistence of symmetry-protected edge states with a gapless bulk gives rise to intriguing topological phenomena absent in gapped counterparts, including nontrivial conformal boundary conditions~\cite{Yu2022PRL,Parker2018PRB}, algebraically localized edge modes~\cite{Verresen2021PRX,Yang2025CP}, universal bulk-boundary correspondence~\cite{Yu2022PRL,Zhang2024PRA,Zhong2025PRB,guo2025generalizedlihaldanecorrespondencecritical}, and intrinsically gapless topological phases~\cite{Thorngren2021PRB,yang2025deconfinedcriticalityintrinsicallygapless}.

A crucial open question is how the concept of symmetry-enriched quantum
criticality manifests in non-Hermitian systems. More importantly, whether non-Hermitian systems host novel symmetry-enriched topological phenomena beyond Hermitian counterparts? If so, how can the underlying mechanisms behind these phenomena be theoretically understood? Progress on this front remains challenging because studying generic non-Hermitian many-body systems is notoriously difficult, plagued by numerical instabilities and inaccuracies that obscure the underlying physics~\cite{Hayata2021PRB,Yamamoto2022PRB,Shen2023PRL,Yu2024PRL_b,Zhong2025PRL, Trefethen2005}. 

In this Letter, 
we systematically explore $\mathcal{PT}$-symmetry–enriched non-unitary criticality in a family of one-dimensional non-Hermitian free-fermion models, uncovering a new class of non-unitary critical points that host robust topological edge modes. Using the combination of exact solution and numerical simulations, we show that although all critical points in these models are described by non-unitary CFTs with central charge $c=-2$, some of them can be further enriched by $\mathcal{PT}$ symmetry, giving rise to topologically non-unitary criticality with stable edge states. Remarkably, the associated topological degeneracies are protected by $\mathcal{PT}$ symmetry and can emerge in critical non-Hermitian free-fermion models. 
We find that, at this $\mathcal{PT}$-symmetry-enriched quantum criticality, the $c=-2$ conformal scaling of the entanglement entropy necessarily comes with a quantized imaginary subleading term, whose quantization is set by the number of entangling boundary modes in the reduced density matrix.
We numerically validate that these entangling boundary modes are robust against $\mathcal{PT}$-symmetric disorder and interactions. These entangling boundary modes are the localized states in the entanglement Hamiltonian and their degeneracy is exactly what the the Affleck-Ludwig $g$-factor measures in the partition~\cite{PhysRevLett.67.161}.
Finally, we elucidate the emergence of topological edge states at this critical point by proposing an entirely new mechanism—generalized mass inversion—that is absent in Hermitian counterparts.

\section{Model}
\begin{figure}[htbp]
  \centering
  \begin{minipage}[b]{1.0\linewidth}
    \panel{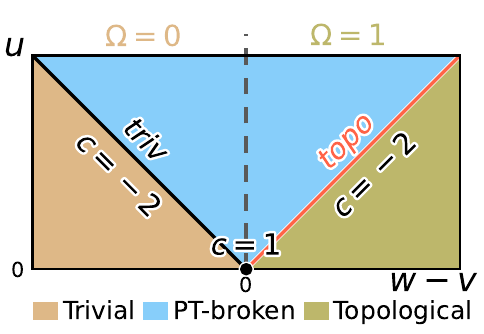}{\linewidth}{(a)}
  \end{minipage}
  \begin{minipage}[b]{0.5\linewidth}
    \panel{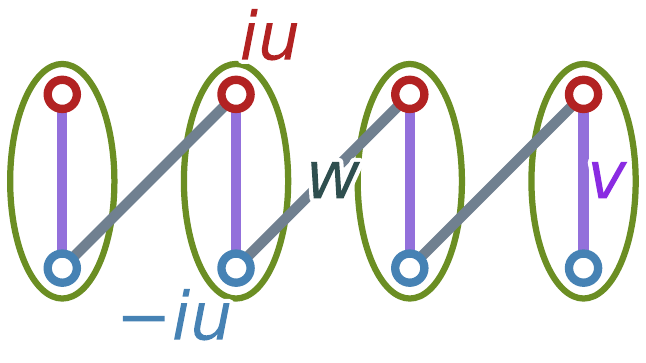}{\linewidth}{(b)}
  \end{minipage}\hfill
  \begin{minipage}[b]{0.48\linewidth}
    \panel{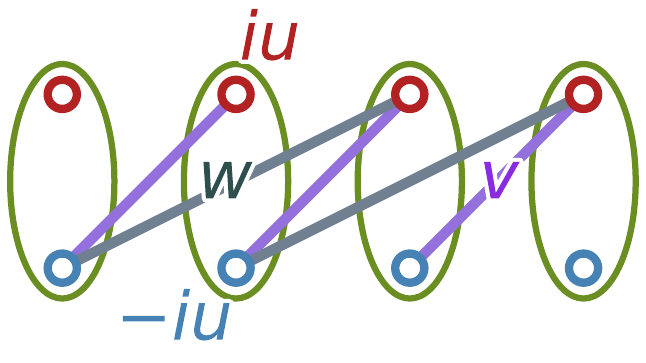}{\linewidth}{(c)}
  \end{minipage}
  \caption{(a) Phase diagram of the $\mathcal{PT}$-symmetric non-Hermitian SSH models in the $(u,w-v)$ plane.
  The Hermitian critical point at $(0,0)$ splits into two non-Hermitian critical lines $u=\pm(w-v)$ that border the $\mathcal{PT}$-broken region.
  The dashed line $w=v$ separates winding sectors $\Omega=0$ and $\Omega=1$.
  (b) Schematic non-Hermitian SSH chain: red/blue circles denote $\pm iu$; black/purple bonds denote Hermitian hoppings.
  (c) Schematic $\alpha=2$ generalization.}
  \label{fig:phase}
\end{figure}

We consider the non-Hermitian Su–Schrieffer–Heeger (SSH) chain with a staggered imaginary on-site potential. In momentum space, the Bloch Hamiltonian is
\begin{equation}
\mathcal{H}_k=
\begin{pmatrix}
 i u & v_k\\
 v_k^{*} & - i u
\end{pmatrix},
\qquad
v_k = v - w e^{-ik},
\label{eq:Hk}
\end{equation}
where $k$ is the single-particle momentum,  $v,w\in\mathbb{R}_{+}$ are the intra-/inter-cell hoppings, and $u\in\mathbb{R}_{+}$ is the staggered imaginary on-site potential (see Fig.~\ref{fig:phase}(b) for a schematic). With spatial-inversion parity symmetry $\mathcal P$ and time reversal symmetry $\mathcal T=\mathcal K$, the Bloch Hamiltonian satisfies $\sigma_x \mathcal{H}_k^* \sigma_x = \mathcal{H}_k$, i.e., the model is $\mathcal{PT}$-symmetric.
The bulk dispersion $E_k=\pm\sqrt{|v_k|^2-u^2}$
is real when $\min_k|v_k|\geq u$ ($\mathcal{PT}$-symmetric) and becomes complex otherwise, signaling spontaneous $\mathcal{PT}$ symmetry-breaking.

Throughout this work, we employ the biorthogonal basis (see the Supplemental Material (SM)~\cite{SM} for a briefly review) and define the many-body density matrix as $\rho=\ket{G_R}\bra{G_L}$, where $\ket{G_R} (\ket{G_L})$ denotes the many-body ground state of $\mathcal{H} (\mathcal{H}^\dagger)$, normalized such that $\langle G_L|G_R\rangle=1$.

The global phase diagram of the non-Hermitian SSH model is schematically shown in Fig.~\ref{fig:phase} (a). It consists of a $\mathcal{PT}$-broken region and two $\mathcal{PT}$-symmetric regions (trivial and topological), distinguished by the winding number $\Omega$ defined as
$\Omega=\frac{1}{2\pi}\int_{-\pi}^{\pi}\! d k\,\partial_k\arg v_k.$
At $u=0$, the Hermitian SSH chain has a single quantum critical point (QCP) at $w=v$ (black dot).
For $u>0$, this point splits into two straight critical lines $u=\pm(w-v)$ that border the $\mathcal{PT}$-broken wedge.
The left critical line lies in the $\Omega=0$ sector and is trivial, whereas the right critical line lies in $\Omega=1$ and is topological in the sense that it supports edge modes along a gapless phase boundary, which is the central focus of this work.
The vertical dashed line $w=v$ marks where the winding number—and hence the number of localized edge modes under open boundary condition (OBC)—changes. Notably, inside the $\mathcal{PT}$-broken wedge, this dashed line is not a phase boundary, since the Berry (Zak) phase that determines phase boundaries no longer coincides with the winding number there~\cite{PhysRevA.87.012118,Lieu2018PRB,SM}. Richer topological features can be accessed by considering the $\alpha$-extended non-Hermitian SSH model, which introduces long-range hopping in the form $v_k^{(\alpha)} = v e^{-i(\alpha-1)k} - w e^{-i\alpha k}$ (see Fig.~\ref{fig:phase}(c) for $\alpha=2$). Remarkably, the overall structure of the phase diagram remains unchanged under this extension, while the long-range hopping increases the number of topological edge modes supported along the critical lines.

\emph{Symmetry-enriched non-unitary CFTs}---We now show that the topological critical line of the non-Hermitian SSH chain realizes a non-unitary CFT with central charge \(c = -2\). Our main probe is the bipartite entanglement entropy under periodic boundary condition (PBC). Using the correlation-matrix method~\cite{Peschel_2009,SM}, the entanglement entropy is computed as
$S_A(\ell_A)
 = -\sum_n \bigl[ \nu_n \log \nu_n + (1 - \nu_n)\log(1 - \nu_n) \bigr]$,
where $\nu_n$ are the eigenvalues of the biorthogonal correlation matrix
$\langle G_L | c_i^\dagger c_j | G_R \rangle$ for a subsystem of length $\ell_A$.

\begin{figure}[htbp]{
  \centering
  \begin{minipage}[b]{0.5\linewidth}
    \panel{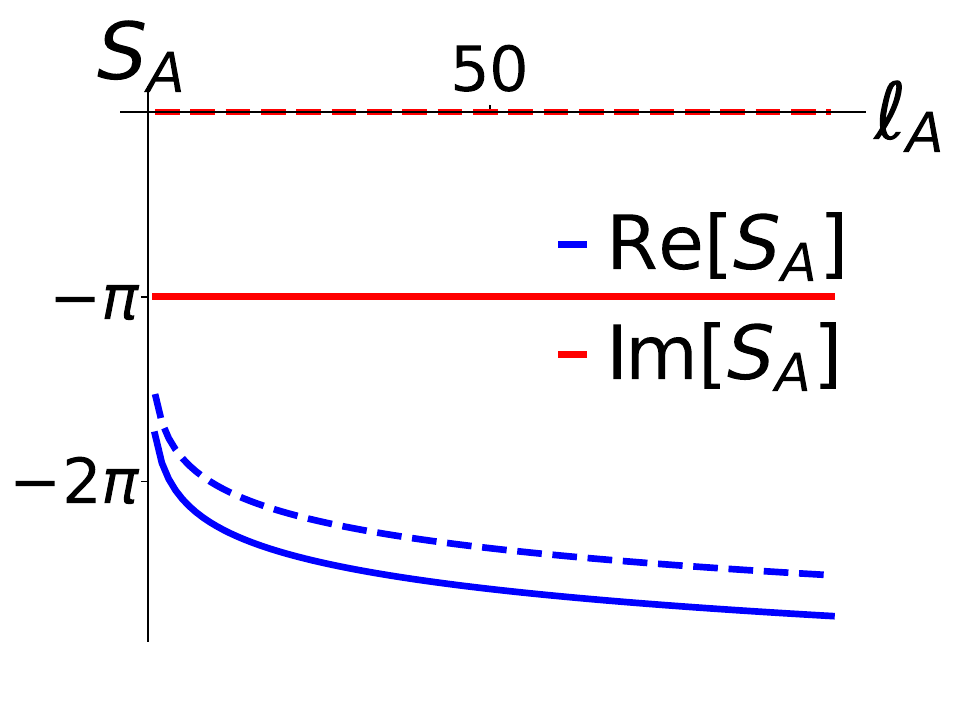}{\linewidth}{(a)}
  \end{minipage}\hfill
  \begin{minipage}[b]{0.5\linewidth}
    \panel{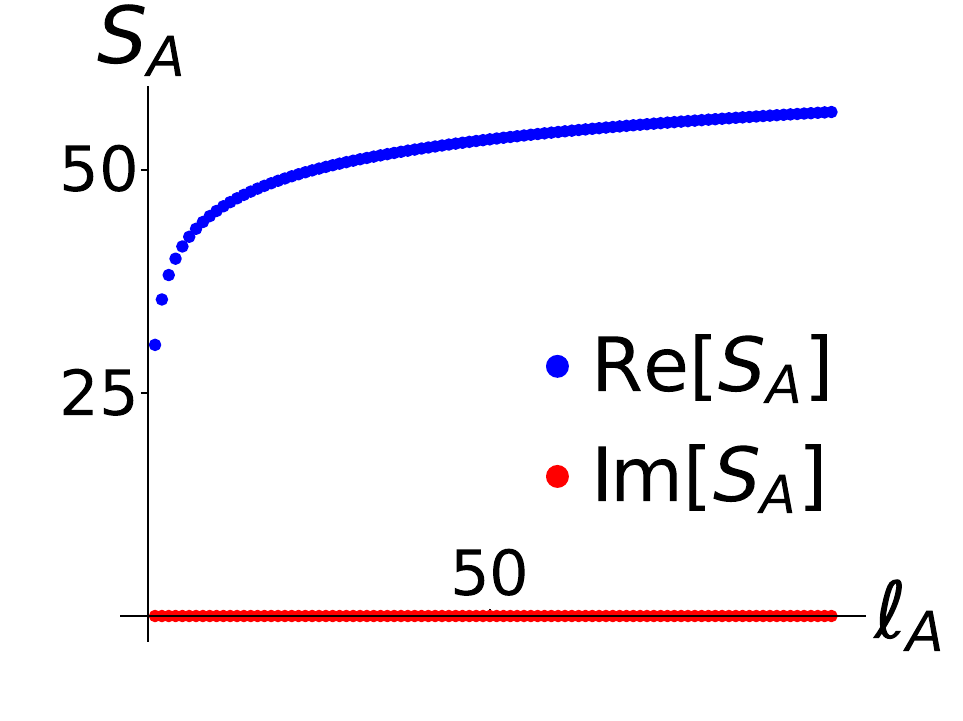}{\linewidth}{(b)}
  \end{minipage}\\[0.8ex]
  \begin{minipage}[b]{0.5\linewidth}
    \panel{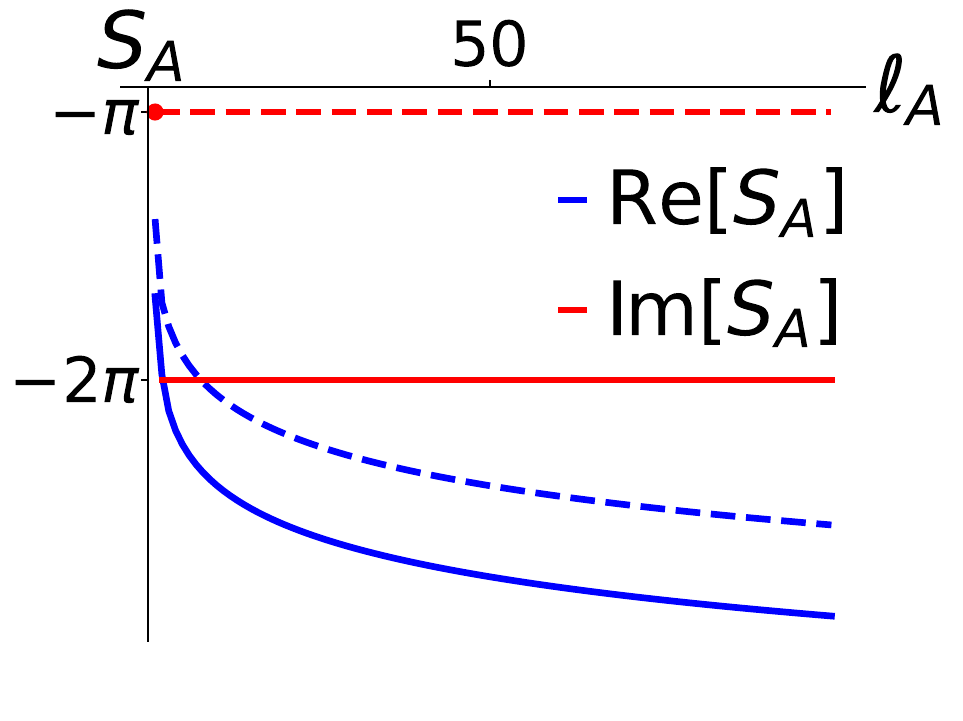}{\linewidth}{(c)}
  \end{minipage}\hfill
  \begin{minipage}[b]{0.5\linewidth}
    \panel{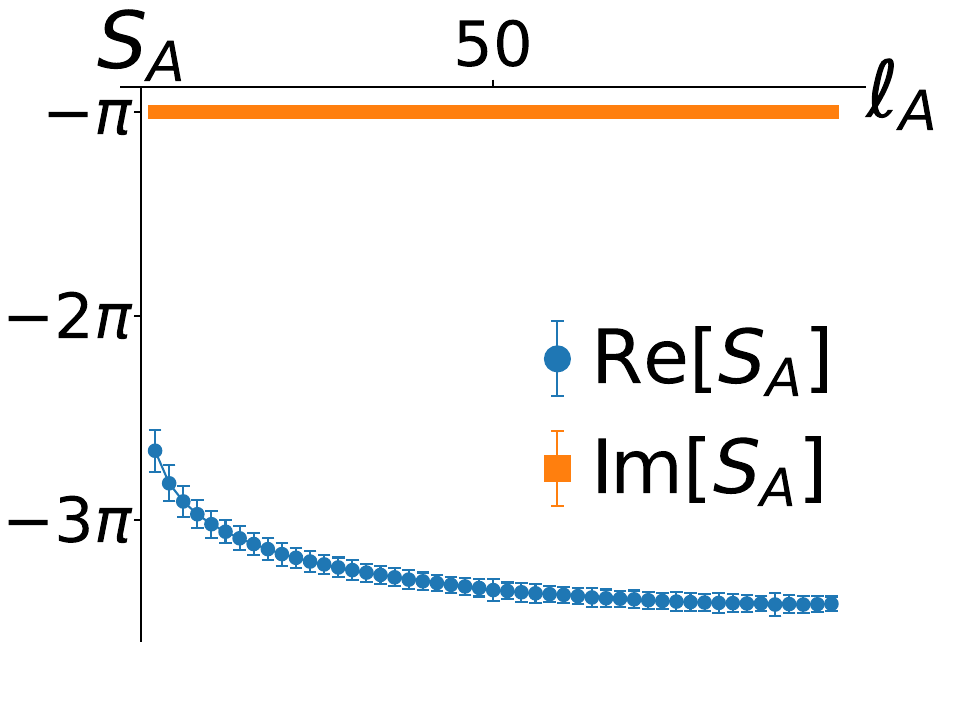}{\linewidth}{(d)} 
  \end{minipage}
   \caption{Entanglement entropy at critical lines of the non-Hermitian SSH family under PBC for a large chain ($L=10000$).
Blue (red) curves show $\Re[S_A]$ ($\Im[S_A]$).
(a) $\alpha=1$ (non-Hermitian SSH) evaluated on the physically (conformally) consistent branch: dashed (solid) curves correspond to the trivial (topological) QCP. In both cases, $\Re[S_A(\ell_A)]$ follows the same $c=-2$ conformal scaling, while the topological QCP carries a quantized constant imaginary part.
(b) The same $\alpha=1$ topological QCP evaluated on the principal branch of the logarithm, where $S_A(\ell_A)$ deviates strongly from the conformal form and bends upward.
(c) $\alpha=2$ extension: dashed (solid) curves denote the QCP with lower (higher) winding number. The real part remains consistent with the $c=-2$ conformal scaling, whereas the imaginary part is quantized and tracks the winding number $\Omega$, illustrating Eq.~(5).
(d) Disorder-averaged entanglement entropy $S_A$ at the topological QCP for a chain of length $L=200$ with $v(x)=-1-\delta_1(x)$, $w(x)=3+\delta_2(x)$, and $u(x)=2-\delta_1(x)+\delta_2(x)-10^{-10}$. The disorder satisfies inversion symmetry, $\delta_{1,2}(x)=\delta_{1,2}(-x)$, with $\delta_{1,2}(x)\in[-0.9,0.9]$ (uniform). Averaged over 1000 realizations; error bars are $\pm1$ s.e.m. The imaginary part is identical for all realizations. A fit to $\mathrm{Re}\,S_A$ yields $c/3=-0.6683$.}
  \label{fig:critical}}
\end{figure}

At the trivial critical line of the non-Hermitian SSH model, due to the biorthogonal setup the spectrum $\{\nu_n\}$ contains eigenvalues with $\nu_n>1$ or $\nu_n<0$, so that the corresponding entanglement entropy can become negative. Nevertheless, the spectrum still organizes into pairs $(\nu,1-\nu)$, and guarantees that $S_A$ remains real. As shown in previous work~\cite{Chang2020PRR} and reproduced in Fig.~\ref{fig:critical} (a), the entanglement entropy obeys the conformal scaling
with central charge $c = -2$ and showing that the biorthogonal construction faithfully captures the physics of a non-unitary CFT.

The topological critical line is more subtle. Here the correlation spectrum acquires a single complex-conjugate pair $\nu_\pm = \frac{1}{2} \pm i I$, whose imaginary part $I$ varies with subsystem size $\ell_A$. This complex pair corresponds to an entangling edge mode that mimics the physical edge state under OBC~\cite{SM}. This complex pair $\nu_\pm$ causes ambiguity due to the multivalue of the logarithm of a complex number,
$\log z = \ln|z| + i(\phi + 2\pi m), m \in \mathbb{Z}$,
where $\phi = \arg(z)$ is the principal argument. For a complex eigenvalue $\nu_n$, the extra phase $i 2\pi m$ enters the factor $\nu_n \log \nu_n$ and can modify even its real part. Consequently, the contribution of $\nu_\pm$ to $S_A$ depends on the choice of branch for $\log z$.

In previous work~\cite{Chang2020PRR}, the edge-mode contribution was evaluated on the principal branch, yielding a real value with a large positive term that grows with \(|I|\). As a result, the total entropy increases with \(\ell_A\) and bends upward, deviating strongly from the conformal scaling of the trivial critical line, as illustrated in Fig.~\ref{fig:critical} (b). However, the trivial and topological critical lines should be governed by the same bulk low-energy field theory, so it is clearly inconsistent that they exhibit different entanglement entropy scaling.

The resolution is to relax the constraint of having a real $S_A$ and choose a branch in which one of the logarithms of the conjugate pair $\nu_\pm$ is shifted by $2\pi i$ from the principal value~\cite{SM}. This choice of branch removes the unphysical linear-in-$\lvert I\rvert$ contribution while leaving an additional imaginary term $-i\pi$, so that the entanglement entropy along the topological critical line takes the form
$S_A(\ell_A)
 = -\frac{2}{3}\log \ell_A - i\pi + \text{const.}$,
with the expected $c=-2$ scaling as shown in Fig.~\ref{fig:critical} (a). In this sense, the requirement of conformal scaling forces the entanglement entropy to carry an imaginary constant $-i\pi$,
which can be interpreted as 
the subleading contribution from the Affleck-Ludwig $g$-factors of the underlining CFTs. I.e., $S_A(\ell_A)
 = \frac{c}{3}\log \ell_A +\ln g_a + \ln g_b$, where $\ln g_{a/b}$ is the $g$-factor associated with the conformal boundary conditions at the endpoint ($a/b$)~\cite{PhysRevLett.67.161,Ohmori_2015,SM}.

The extension to the $\alpha$-extended non-Hermitian SSH models is straightforward. At a topological critical line within the $\Omega$ sector, the entanglement spectrum contains $\Omega$ pairs of entanglement edge modes with \(\nu_a = \tfrac{1}{2} \pm i I_a\). The requirement of conformal scaling fixes the branch choice for each pair and yields 
\begin{equation}
    S_A(\ell_A) = -\frac{2}{3}\log \ell_A - i\pi\,\Omega + \mathrm{const.}
\end{equation}
Thus, across all topological critical lines of the $\alpha$–extended non-Hermitian SSH model, the real part of the entropy realizes a non-unitary CFT with $c=-2$, while the quantized imaginary part simply counts how many pairs of entanglement edge mode are present, i.e., the winding number $\Omega$ [see Fig.~\ref{fig:critical} (c)].

The imaginary constant entanglement entropy originated from the entangling boundary modes is robust under the symmetry-preserving disorder while keeping the system at criticality. As shown in Fig.~\ref{fig:critical} (d), the real part of the entanglement entropy continues to obey conformal scaling with central charge $c = -2$,
while the imaginary part of the entanglement entropy is set by the winding number—remains pinned to the same value across all disorder realizations.
We also consider the interaction effect on the robustness of the entangling boundary modes. By adding the nearest-neighbor Hubbard interaction $U$ to the non-Hermitian SSH and mapping it to the staggered XXZ chain via the Jordan–Wigner transformation, the entangling edge modes in the reduced density matrix are present and contributed to the quantized imaginary entanglement entropy~\cite{SM}. In addition, subleading terms of the R\'enyi entropies $S^{n}_A(\ell_a)$ with $n$ being odd integers also exhibit the quantized imaginary entanglement entropy~\cite{SM}.

A complementary signature of criticality comes from the finite-size scaling of the ground-state energy under PBC. In CFT, this takes the form
$E_0^{\mathrm{PBC}}(L) = L \epsilon - \frac{\pi v_F c}{6L} + \cdots$,
where $\epsilon$ is the bulk energy density and $v_F$ is the Fermi velocity. For all $\alpha$-extended non-Hermitian SSH models, the bulk dispersion remains identical across phases, enabling a unified scaling analysis for all critical lines. As shown in Fig.~\ref{fig:mass_inv} (a), the coefficient of the $1/L$ term matches the CFT prediction with $c=-2$, corroborating the entanglement-entropy results~\cite{SM}.


\section{$\mathcal{PT}$ symmetry-protected topological edge states at non-unitary criticality.}
We have identified non-unitary critical lines in the $\alpha$-extension non-Hermitian SSH model that reside in distinct topological sectors. We now demonstrate that these critical lines cannot be smoothly connected without breaking $\mathcal{PT}$ symmetry, implying that the physical edge modes at the topological critical line are protected by $\mathcal{PT}$ symmetry. 
From the dispersion $E_k = \pm \sqrt{|v_k|^2 - u^2}$, the $\mathcal{PT}$-symmetric regime is defined by the condition $|v_k| \ge u$ for all $k$, so that the spectrum remains real. In particular, this enforces $v_k \neq 0$ throughout the Brillouin zone, so the loop traced by $v_k$ in the complex plane never crosses the origin and the winding number is pinned. Once the condition $\min_k |v_k| \ge u$ is violated, the system enters the $\mathcal{PT}$-broken regime with a complex spectrum. 
In this region, $v_k$ can pass through zero without any additional gap-closing or reopening~\cite{Yuce2018}, so the winding number may change continuously without a phase transition. Consequently, topological edge modes are unstable in the $\mathcal{PT}$-broken region, even though the winding number may remain finite in this case. This is precisely why the dashed line $w = v$ in Fig.~\ref{fig:phase} (a) does not represent a true phase boundary inside the $\mathcal{PT}$-broken wedge. The protection is also lost under explicit $\mathcal{PT}$ symmetry breaking, which generically produces a complex gap~\footnote{For example, adding a real staggered on-site potential $m$ shifts the diagonal to $(m + i u)\sigma_z$, giving $E^2(k)=|v_k|^2 + (m + i u)^2 = |v_k|^2 + m^2 - u^2 + 2 i m u$, there will be a nonzero complex gap for any $m \neq 0$ and $u \neq 0$. Thus one can tune $v_k$ through zero and change $\Omega$ without gap closing.}. In this context, the spectrum can remain gapped while $v_k$ crosses zero, again allowing the winding number to change without a phase transition.



\begin{figure}[htbp]{
  \centering
  \begin{minipage}[b]{0.518\linewidth}
    \panel{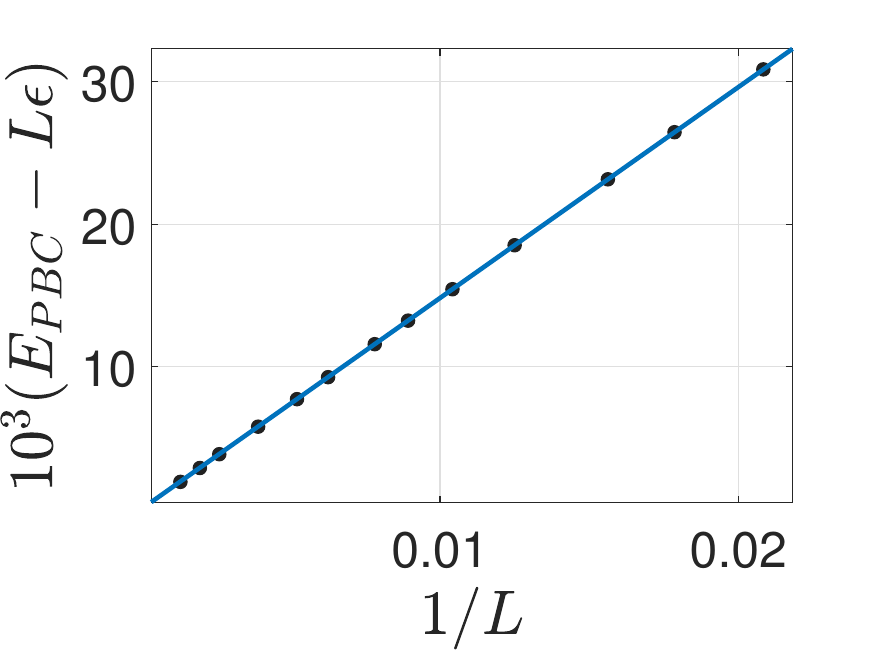}{\linewidth}{(a)}
  \end{minipage}\hfill
  \begin{minipage}[b]{0.482\linewidth}
    \panel{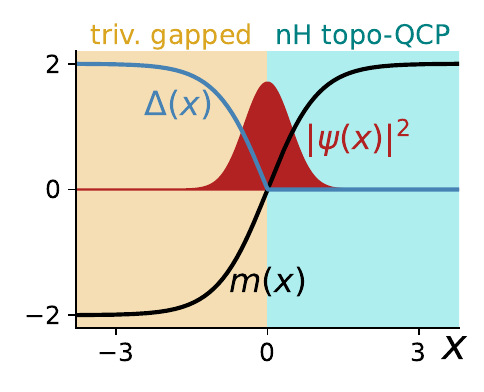}{\linewidth}{(b)}
  \end{minipage}
  \caption{(a) Finite-size scaling of the ground-state energy under PBC, consistent with the non-unitary CFT description with $c=-2$ (Fermi velocity $v_F=\sqrt{2}$ here). (b) Interface realizing generalized mass inversion between a trivial gapped region (left) and the non-Hermitian SSH chain at its topological QCP (right). Shown are the spatial profiles of the real mass $m(x)$ (black), the effective bulk gap $\Delta(x)=\sqrt{m(x)^2-u(x)^2}$ (blue), and the bound-state density $|\psi(x)|^2$ (red) localized at the interface. In this geometry neither mass component is uniform at the interface, so the eigenvalue equations remain coupled; an explicit continuum (field-theory) solution demonstrating that generalized mass inversion still pins an edge state at the interface is presented in the SM~\cite{SM}.}
  \label{fig:mass_inv}}
\end{figure}

\emph{Generalized mass inversion in $\mathcal{PT}$ symmetry-enriched non-Hermitian QCPs}—
In Hermitian free-fermion systems, symmetry-enriched quantum criticality can only be realized by long-range hopping in the $\alpha\!\ge\!2$ chain~\cite{Verresen2018PRL}, through the “kinetic inversion” mechanism~\cite{verresen2020topologyedgestatessurvive}. By contrast, in the non-Hermitian SSH model, we can have topological edge modes survive at criticality without long-range hopping, due to a generalization of mass inversion. 

To elucidate this new mechanism, we begin by solving the edge mode analytically for the non-Hermitian SSH model. The low-energy linearized Hamiltonian of the model can be written as
\begin{equation}
    \mathcal{H} =
    \begin{bmatrix}
        iu & -\partial_x + m \\
        \partial_x + m & -iu
    \end{bmatrix},
\end{equation}
where $m \sim w - v$ is the effective mass set by the difference between intra-cell and inter-cell hoppings, and $u$ is the non-Hermitian on-site potential. The eigenequations
\begin{align}
(-\partial_x + m)\psi_B(x) &= (E - iu)\psi_A(x), \nonumber\\
(\partial_x + m)\psi_A(x) &= (E + iu)\psi_B(x)
\end{align}
where $A/B$ label the two sublattices and $\psi(x) = \big(\psi_A(x), \psi_B(x)\big)^{T}$. These equations decouple at $E = \pm iu$: for $E = iu$ we can set $\psi_B = 0$, so the second line reduces to
\begin{equation}
    (\partial_x + m)\psi_A(x) = 0 \quad \Rightarrow \quad \psi_A(x) \sim e^{-mx},
\end{equation}
which is a normalizable edge solution for the $x>0$ region with $m>0$. Thus we see that an edge mode remains localized at the boundary even as the bulk gap $\Delta = \sqrt{m^2 - u^2}$ closes. (see SM~\cite{SM} for lattice derivation.)

This analysis can be extended to a more general setup. We now consider a Dirac Hamiltonian with multiple mass terms, $M = \sum_a m_a \sigma_a$, and an interface where one particular mass component changes sign across the interface. Since all Pauli matrices are equivalent up to a basis rotation, we can choose a basis in which the mass component that remains uniform across the interface, $m_{\text{uniform}}$, lies in the $\sigma_z$ sector. In that basis, choosing the energy $E = \pm m_{\text{uniform}}$ again decouples the eigenequations, yielding an interface-localized mode (see Fig.~\ref{fig:mass_inv} (b) for a schematic) whose decay rate $\kappa$ is fixed by the sign-changing mass component $m_{\text{bdy}}$, i.e., $\kappa = |m_{\text{bdy}}|$. We thus see that two physical scales naturally decouple when multiple mass terms are present: (i) the decay rate $\kappa$, set by the mass component that flips sign across the boundary ($m_{\text{bdy}}$); and (ii) the bulk gap $\Delta$, set by the total mass, $\Delta = \sqrt{\sum_a m_a^{2}}$. We refer to this separation of scales, together with ordinary mass inversion in topological insulator~\cite{Jackiw1976}, as \emph{generalized mass inversion}.

Within this mechanism, we can now reinterpret why the non-Hermitian SSH model realizes a QCP that hosts topological edge modes. For the non-Hermitian SSH model, the imaginary mass $iu$ plays the role of the uniform mass, while the sign-changing component is the real mass $m$, so the decay rate is $\kappa = m$ and the bulk gap is $\Delta = \sqrt{m^{2}-u^{2}}$. The cancellation between mass components allows us to tune $u \to m$ to close the bulk gap $(\Delta \to 0)$ while keeping the decay rate $\kappa = m$ finite, hence realizing a symmetry-enriched quantum
criticality. In contrast, in Hermitian Dirac models, all mass components are real and contribute positively to $\Delta$, so the gap cannot close without driving the entire mass vector to zero and delocalizing the edge mode.

\section{Discussion and concluding remarks.}
To summarize, we uncover a new class of non-unitary criticality enriched by $\mathcal{PT}$ symmetry that hosts robust topological edge states. Specifically, we construct a broad family of one-dimensional $\mathcal{PT}$-symmetric non-Hermitian free-fermion models with $\alpha$-range hopping, all of which exhibit QCPs described by non-unitary CFTs with central charge $c=-2$, as numerically confirmed by the scaling of the entanglement entropy and the ground-state energy. More importantly, we demonstrate unambiguously that these non-Hermitian critical points are topologically distinct, in the sense that they cannot be smoothly connected without either breaking $\mathcal{PT}$ symmetry or encountering a multicritical point. 
 Remarkably, we find that the subleading term of entanglement entropy is purely imaginary and quantized to the number of entangling boundary modes in the reduced density matrix, which can be interpreted as the Affleck-Ludwig $g$-factor associated with this symmetry-enriched non-unitary CFT.
Finally, we demonstrate that these edge modes are enforced by a new mechanism—generalized mass inversion—which fundamentally distinct from that in Hermitian counterparts.

Looking ahead
to the experimental realization, non-Hermitian quantum systems can be simulated on photonic platforms, where gain and loss are experimentally implemented~\cite{Guo2009, Ruter2010, Peng2014, Chang2014, Lin2024MeasureNC}. 
Furthermore, the entanglement properties reported in this work could potentially be accessed by embedding the Hamiltonian into an enlarged Hermitian system with ancillas and extracting the required biorthogonal correlators through joint measurements~\cite{Matsumoto2022}, and the quantum-circuit protocol which the biorthogonal winding numbers for the non-Hermitian SSH model were extracted~\cite{PhysRevA.107.052205}. 

\textit{Note added:} After completing this manuscript, we became aware of a related independent study on the critical edge states in one-dimensional non-Hermitian free-fermion chains~\cite{zhou2025topological}.

\section*{Acknowledgements}
We thank Xueda Wen for helpful discussion. P.-Y.C acknowledges support from RIKEN Center for Interdisciplinary Theoretical and Mathematical Sciences and National Center for Theoretical Sciences, Physics Division. This research was supported in part by Perimeter Institute for Theoretical Physics. Research at Perimeter
Institute is supported by the Government of Canada through the Department of Innovation, Science, and
Economic Development, and by the Province of Ontario through the Ministry of Colleges and Universities.


\paragraph{Funding information}
K.-H. Chou and P.-Y. Chang was supported by National Science and Technology Council of Taiwan under Grants No. NSTC 113-2112-M- 007-019, 114-2918-I-007-015.
X.-J. Yu was supported by the National Natural Science Foundation of China (Grant No.12405034) and a start-up grant from Fuzhou University and Eastern Institute of Technology, Ningbo.

\begin{appendix}

\section{Biorthogonal (Right--Left) framework}
\label{sm1}

Non-Hermitian Hamiltonians $H^\dagger\neq H$ generically admit distinct right and left eigenstates, so specifying the ``state'' of the system requires fixing a left--right convention. In this work we adopt the biorthogonal (right--left, RL) formulation. We introduce right and left eigenstates satisfying
\begin{equation}
H\ket{\Psi_{R,\alpha}} = E_\alpha\ket{\Psi_{R,\alpha}},
\qquad
\bra{\Psi_{L,\alpha}}H = E_\alpha\bra{\Psi_{L,\alpha}},
\end{equation}
together with the biorthonormal relation
\begin{equation}
\braket{\Psi_{L,\alpha}}{\Psi_{R,\beta}}=\delta_{\alpha\beta}.
\end{equation}
This replaces the usual Hermitian orthonormality and provides a consistent normalization (and completeness) structure for non-Hermitian eigenbases.

Within the RL framework, a single eigenstate is naturally encoded by the biorthogonal density operator~\cite{Brody_2014}
\begin{equation}
\rho_{RL}\equiv \ket{\Psi_R}\bra{\Psi_L}.
\end{equation}
For the many-body ground state, we denote the eigenpair by $\ket{G_R}$ and $\bra{G_L}$, and fix the overall normalization by $\langle G_L|G_R\rangle=1$, so that ${\rm Tr}(\rho_{RL})=1$. Static observables are computed in the standard density-matrix manner,
\begin{equation}
\langle O\rangle \equiv {\rm Tr}\left(\rho_{RL} O\right) = \bra{\Psi_L}O\ket{\Psi_R}.
\end{equation}

For a spatial bipartition $A\cup\bar A$, we define the reduced operator by the partial trace over $\bar A$,
\begin{equation}
\rho_A={\rm Tr}_{\bar A}\,\rho_{RL},
\end{equation}
and evaluate the entanglement entropy from $\rho_A$ via the von Neumann functional
\begin{equation}
S_A=-{\rm Tr}\big(\rho_A\ln\rho_A\big).
\end{equation}

To formulate the discussion purely at the level of eigenstates---and in particular to keep $\bra{\Psi_L}$ an eigenbra under time evolution---we adopt the adjoint action of $H$ on both sides~\cite{Mannheim2013PT},
\begin{equation}
\rho_{RL}(t)=e^{-iHt}\,\rho_{RL}(0)\,e^{+iHt}.
\end{equation}
This choice makes the RL density operator strictly stationary for an energy eigenstate. Concretely, if $\rho_{RL}(0)=\ket{\Psi_R}\bra{\Psi_L}$ corresponds to a single eigenvalue $E$ (which may be complex), then
\begin{equation}
\rho_{RL}(t)=e^{-iEt}\ket{\Psi_R}\bra{\Psi_L}e^{+iEt}=\rho_{RL}(0).
\end{equation}
As a result, ${\rm Tr}(\rho_{RL})$ is time independent, and all observables computed from $\rho_{RL}$ remain unchanged in time. The RL formulation therefore provides a stationary notion of an eigenstate and a natural analogue of a closed-system eigenstate for a non-Hermitian Hamiltonian.

It is useful to contrast this with the right--right (RR) prescription,
\begin{equation}
\rho_{RR}=\ket{\Psi_R}\bra{\Psi_R},
\end{equation}
whose non-unitary evolution takes the form
\begin{equation}
\rho_{RR}(t)=e^{-iHt}\,\rho_{RR}(0)\,e^{+iH^\dagger t}.
\end{equation}
When the spectrum contains complex eigenvalues, the norm ${\rm Tr}[\rho_{RR}(t)]$ is generally not conserved, and the evolution naturally admits an interpretation in terms of probability loss or gain to an external environment. In that setting one often works with a continuously re-normalized density matrix $\tilde\rho_{RR}(t)=\rho_{RR}(t)/{\rm Tr}[\rho_{RR}(t)]$, which remains positive and is well suited to open-system physics such as decoherence, relaxation, and steady states.

Although these two prescriptions may start from the same non-Hermitian Hamiltonian $H$, they are physically distinct because they employ different density operators and different time-evolution conventions. Consequently, they can yield different expectation values, different entanglement behavior, and even different phase diagrams, and should not be conflated.

Finally, because $\rho_{RL}$ (and hence $\rho_A$) is generically non-Hermitian, the resulting entropy is not constrained to be positive and may in general be complex. This feature allows the biorthogonal (RL) formulation to naturally accommodate non-unitary CFT behavior, including negative or complex central charges, as demonstrated in previous work~\cite{Couvreur2017, Chang2020PRR,Tu2022SciPost, Lee2022PRL, Chang2023SciPostCore, Rottoli_2024, Xue2026, Li_PRB,linden2025, n578-ljd5}.

\section{The relation between Berry phase and winding number in non-Hermitian systems}
\label{sm2}

In this section, we review the complex Berry phase, following Refs.~\cite{PhysRevA.87.012118,Lieu2018PRB}. Recall the non-Hermitian SSH Bloch Hamiltonian
\begin{equation}
H(k)=\begin{pmatrix} iu & v_k \\ v_k^{*} & -iu \end{pmatrix}
\end{equation}
with $u\in\mathbb{R}$. The lower band eigenstate can be express as
\begin{equation}
\label{eqS:eigs}\tag{S1}
\ket{R_{k,-}}=
\begin{bmatrix}
- e^{-i\varphi(k)}\sin\frac{\theta(k)}{2}\\
\cos\frac{\theta(k)}{2}
\end{bmatrix},
\qquad
\ket{L_{k,-}}=
\begin{bmatrix}
- e^{+i\varphi(k)}\sin\frac{\theta(k)}{2}\\
\cos\frac{\theta(k)}{2}
\end{bmatrix},
\end{equation}
where $\varphi(k)\equiv {\rm arg} v_k$ and $\theta(k)= \arctan [iu/{\rm abs}{v_k}]$.
The biorthogonal Berry connection and Zak phase are
\begin{equation}
\label{eqS:defs}\tag{S2}
A_{--(k)}=i\bra{L_{k,-}}\partial_k\ket{R_{k,-}},
\qquad
Q_-=\int_{-\pi}^{\pi} dk\, A_{--}(k).
\end{equation}
Substituting \eqref{eqS:eigs} into \eqref{eqS:defs} yields
\begin{equation}
\label{eqS:general}\tag{S3}
A_{--}(k)=-\frac{1}{2}\partial_k\varphi(k)
+\frac{1}{2}\partial_k\varphi(k)\cos\theta(k).
\end{equation}
Using $\cos\theta(k)= iu/\sqrt{|v_k|^{2}-u^{2}}$ we obtain
\begin{equation}
\label{eqS:key}\tag{S4}
A_{--}(k)=-\frac{1}{2}\partial_k\varphi(k)
+\frac{1}{2}\partial_k\varphi(k)
\frac{iu}{\sqrt{|v_k|^{2}-u^{2}}}.
\end{equation}

On the $\mathcal{PT}$-symmetric part of the phase diagram---i.e., wherever the spectrum is real ($E(k)\in\mathbb{R}_{\ge0}$, which includes the critical locus)---the factor $iu/\sqrt{|v_k|^{2}-u^{2}}$ is imaginary across the momentum space. Hence the second term of \eqref{eqS:key} contributes only to the imaginary part, and
\begin{equation}
\label{eqS:ReA}\tag{S5}
\operatorname{Re}[A_{--}](k)=-\frac{1}{2}\partial_k\varphi(k).
\end{equation}
Integrating over the Brillouin zone gives the quantized relation
\begin{equation}
\label{eqS:ReQ=piw}\tag{S6}
\operatorname{Re}[Q_-]=-\frac{1}{2}\int_{-\pi}^{\pi} \partial_k\varphi dk
=\pi \Omega,
\qquad
\Omega=\frac{1}{2\pi}\int_{-\pi}^{\pi} \partial_k\arg v_k dk\in\mathbb{Z}.
\end{equation}
Thus, throughout the $\mathcal{PT}$-symmetric part, the real Zak phase equals $\pi\Omega$, and the system has protected topology.

Inside the \emph{$\mathcal{PT}$-broken wedge} there exists a nonempty set
$I=\{k:\ |v_k|<u\}$ on which $E(k)= i\sqrt{u^{2}-|v_k|^{2}}$ and
\begin{equation}
\label{eqS:realfactor}\tag{S7}
\frac{iu}{\sqrt{|v_k|^{2}-u^{2}}}
=\frac{u}{\sqrt{u^{2}-|v_k|^{2}}}\in\mathbb{R}\ (k\in I).
\end{equation}
The connection then acquires an additional real contribution on $I$,
\begin{equation}
\label{eqS:ReA_broken}\tag{S8}
\mathrm{Re}A_{--}(k)=
-\frac{1}{2}\partial_k\varphi(k)
+\frac{1}{2}\partial_k\varphi(k)
\frac{u}{\sqrt{u^{2}-|v_k|^{2}}}\quad(k\in I),
\end{equation}
and the Zak phase becomes
\begin{equation}
\label{eqS:ReQ_broken}\tag{S9}
\mathrm{Re}Q_-=\pi \Omega
+\frac{1}{2}\int_{I} dk\, \partial_k\varphi(k)
\frac{u}{\sqrt{u^{2}-|v_k|^{2}}}.
\end{equation}
The extra integral is non-quantized, so the Berry phase no longer tracks the winding number; the winding-number (and edge-mode) count persists but is no longer robust in this regime.

\section{Correlation-matrix method for quantum entanglement in non-Hermitian systems}
\label{sm3}
We consider a quadratic (free) fermion system whose ground state is Gaussian. For a bipartition $A\cup\bar A$, the reduced state on $A$ can be written as
\begin{equation}
\rho_A=Z^{-1}e^{-\mathcal{H}_E},\qquad 
\mathcal{H}_E=\sum_{i,j\in A} h^E_{ij}c_i^\dagger c_j .
\end{equation}
The correlation matrix on $A$ is defined in trace form by
\begin{equation}
C_{ij}= \bra{G_L}c_i^\dagger c_j\ket{G_R} = {\rm Tr}\big(\rho_Ac_i^\dagger c_j\big),\qquad i,j\in A .
\end{equation}
Equivalently, as a projector onto the occupied single-particle modes,
\begin{equation}
C_{ij}=\sum_{\alpha\in\mathrm{occ}} L_{\alpha i}\,R^\dagger_{\alpha j}.
\end{equation}

Diagonalize the single-particle entanglement Hamiltonian with a similarity transform
\begin{equation}
h^E = S\mathrm{diag}(\varepsilon_n)S^{-1}.
\end{equation}
The new basis can be expressed as $f_{L}=S\phi S^{-1}$ and $f_{R}^{\dagger}=S\phi^{\dagger}S^{-1}$, so that $\{f_{L,m},f_{R,n}^{\dagger}\}=\delta_{mn}$. In this basis
\begin{equation}
\mathcal H_E=\sum_{n}\varepsilon_n\, f_{R,n}^\dagger f_{L,n},
\qquad
\rho_A=\prod_{n}\frac{e^{-\varepsilon_n f_{R,n}^\dagger f_{L,n}}}{1+e^{-\varepsilon_n}},
\end{equation}
which gives the single-mode occupations
\begin{equation}
\nu_n \equiv \langle {f_{R,n}^\dagger f_{L,n}} \rangle=\frac{1}{1+e^{\varepsilon_n}} .
\end{equation}

The von Neumann entropy can be written as
\begin{align}
S_A 
&= -{\rm Tr}\big(\rho_A\ln\rho_A\big) \nonumber\\
&= -{\rm Tr}\left[
\frac{e^{-h^E}}{1+e^{-h^E}}\ln\frac{e^{-h^E}}{1+e^{-h^E}}
+\frac{1}{1+e^{-h^E}}\ln\frac{1}{1+e^{-h^E}}
\right] \nonumber\\
&= -\sum_{n}\left[\nu_n\ln \nu_n+(1-\nu_n)\ln(1-\nu_n)\right],
\end{align}
which is the expression used in the main text.

\section{Self-consistent branch choice for non-Hermitian entanglement entropy}
\label{sm4}

\subsection{Edge–mode case}
\label{subsec:edge-branch}

\begin{figure}[htbp]
	\centering
	\begin{minipage}[b]{0.432\linewidth}
		\panelH{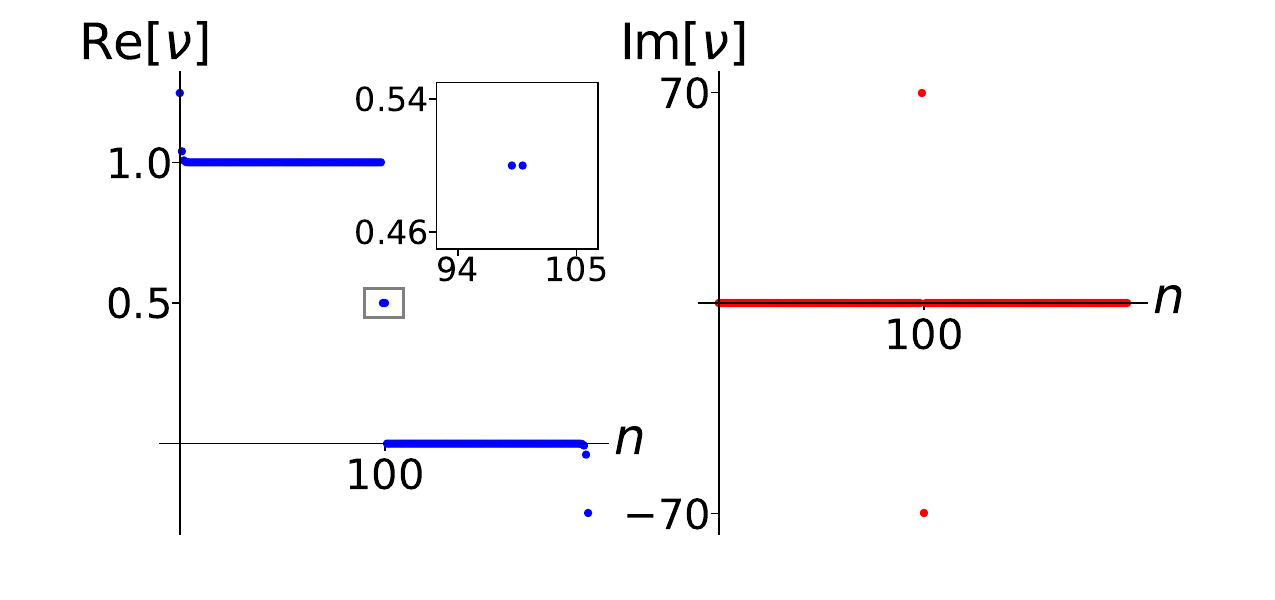}{3.7cm}{(a)}
	\end{minipage}
	\begin{minipage}[b]{0.273\linewidth}
		\panelH{EE_prin-eps-converted-to.pdf}{3.7cm}{(b)}
	\end{minipage}
	\begin{minipage}[b]{0.273\linewidth}
		\panelH{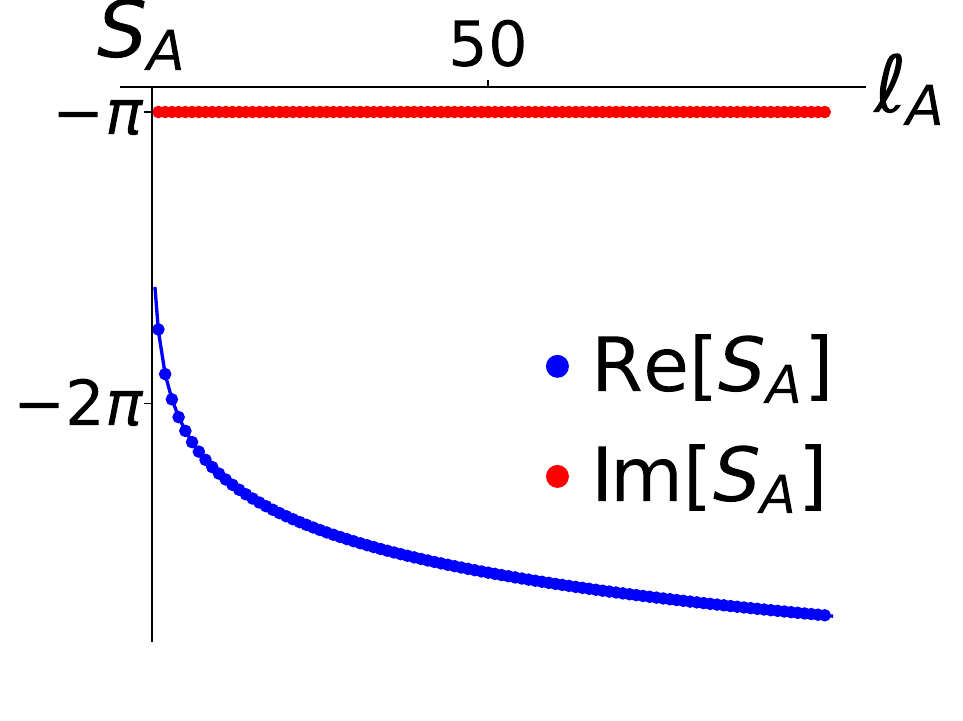}{3.7cm}{(c)}
	\end{minipage}
	\caption{Comparison of branch choices for entanglement in the non-Hermitian SSH chain at its topological QCP
		($L=10^4$; $(v,w,u)=(1,2,1-10^{-12})$).
		(a) Entanglement spectrum $\{\nu\}$ at $\ell_A=100$ (left: $\operatorname{Re}[\nu]$; right: $\operatorname{Im}[\nu]$).
		(b) $S_A(\ell_A)$ evaluated with the principal branch.
		(c) $S_A(\ell_A)$ on the physically consistent branch, exhibiting the expected $c=-2$ scaling.
		Blue (red) markers denote $\operatorname{Re}[S_A]$ ($\operatorname{Im}[S_A]$).}
	\label{fig:edge_cal}
\end{figure}

In nH SSH chains, the biorthogonal correlation spectrum of a subsystem may contain complex eigenvalues. At topological QCPs, one observes a complex-conjugate pair (see Fig.~\ref{fig:edge_cal} (a))
\begin{equation}
\nu_\pm = \frac{1}{2} \pm i\,\mathcal{I}, \qquad \mathcal{I}\in\mathbb{R_+},
\label{eq:nu-pm}
\end{equation}
which contributes to the von Neumann entropy
\begin{equation}
S_A = -\sum_{n} \Big[ \nu_n \ln \nu_n + (1-\nu_n)\ln(1-\nu_n) \Big].
\end{equation}
Note that the logarithm here is a multivalued function: for any nonzero $z=|z|e^{i\phi}$,
\begin{equation}
\ln z = \ln|z| + i\big(\phi + 2\pi m\big), \qquad m\in\mathbb{Z},
\end{equation}
which gives rise to a branch-choice ambiguity in $S_A$ that we resolve in this work.

As stated in the main text, this complex-conjugate pair corresponds to an edge mode. We denote its contribution by
\begin{equation}
S_{\rm edge}
= -2\Big[\big(\frac{1}{2}+i\mathcal{I}\big)\ln\big(\frac{1}{2}+i\mathcal{I}\big)
+ \big(\frac{1}{2}-i\mathcal{I}\big)\ln\big(\frac{1}{2}-i\mathcal{I}\big)\Big].
\label{eq:Sedge-def}
\end{equation}
On the principal branch, with $\mathrm{Arg}\in(-\pi,\pi]$, write
$\frac{1}{2}\pm i\mathcal{I}=r\,e^{\pm i\phi}$, where
$r=\sqrt{\frac{1}{4}+\mathcal{I}^2}$ and $\phi=\arctan(2\mathcal{I})\in(0,\pi)$.
Substituting into \eqref{eq:Sedge-def} yields
\begin{equation}
S_{\rm edge}^{(\rm princ)}
= -2\ln r + 4\phi\mathcal{I},
\label{eq:Sedge-princ}
\end{equation}
which is purely real. However, this choice leads to scaling behavior that deviates markedly from the expected conformal form (see Fig.~\ref{fig:edge_cal} (b)), as pointed out in previous work~\cite{Chang2020PRR}.

From Fig.~\ref{fig:edge_cal} (b), the principal-branch fit yields an entanglement entropy that is positive and increases with subsystem size $\ell_A$, in tension with the expected conformal scaling. Inspecting Eq.~\eqref{eq:Sedge-princ}, we single out the term $4\phi\mathcal{I}$ as a candidate culprit for the positive deviation. At criticality the edge-pair parameter $\mathcal{I}$ is parametrically large (formally $|\mathcal{I}|\to\infty$ if the EP is exactly occupied), so with $\phi=\arctan(2\mathcal{I})$ we have
\begin{equation}
\phi \to \frac{\pi}{2}
\qquad\Rightarrow\qquad
4\phi\mathcal{I} \longrightarrow 2\pi\mathcal{I}.
\end{equation}
This suggestive form motivates a minimal fix. The factor $4\phi$ in front of $\mathcal{I}$ originates from the phase choices of the four logarithmic terms. By shifting the branch of one of these logarithms by $2\pi$ in its argument, we replace $4\phi\mathcal{I}$ with $(4\phi-2\pi)\mathcal{I}$, thereby canceling this positive contribution. With this branch choice, the edge contribution becomes
\begin{equation}
S_{\rm edge}^{(\rm branch)}= -2\ln r + \big(4\phi - 2\pi\big)\mathcal{I} \pm i\pi.
\label{eq:edge-branch}
\end{equation}
We fix the sign convention to the negative one, yielding $\operatorname{Im}[S_{\rm edge}]=-\pi$. Consequently, the entanglement entropy reads
\begin{equation}
S_A = -\frac{2}{3}\ln \ell_A - i\pi + \text{const.}.
\end{equation}
As shown in Fig.~\ref{fig:edge_cal} (c), with this branch reassignment in place, the real-part fits recover the expected conformal scaling. Thus we regard this as the correct, physically meaningful branch choice.

\subsection{Gapped case}

Prior sections clarified that the pathology in the entanglement entropy of the topological nH SSH model is a branch-choice artifact. Earlier studies~\cite{Chang2020PRR} also reported an "entropy flip" at small gaps or under slight boundary deviations, attributing it to boundary or exceptional-point (EP) sensitivity. Re-examining the gapped regime, we find that the apparent entropy flip coincides with the onset of a complex quartet in the correlation spectrum, which we refer to as gapped modes
\begin{equation}
\{\nu,\;\nu^*,\;1-\nu,\;1-\nu^*\}
\end{equation}
in the correlation spectrum. This indicates that the discontinuity is not a genuine phase transition but another branch-choice artifact.

\begin{figure}[htbp]
	\centering
	\begin{minipage}[b]{0.48\linewidth}
		\panelH{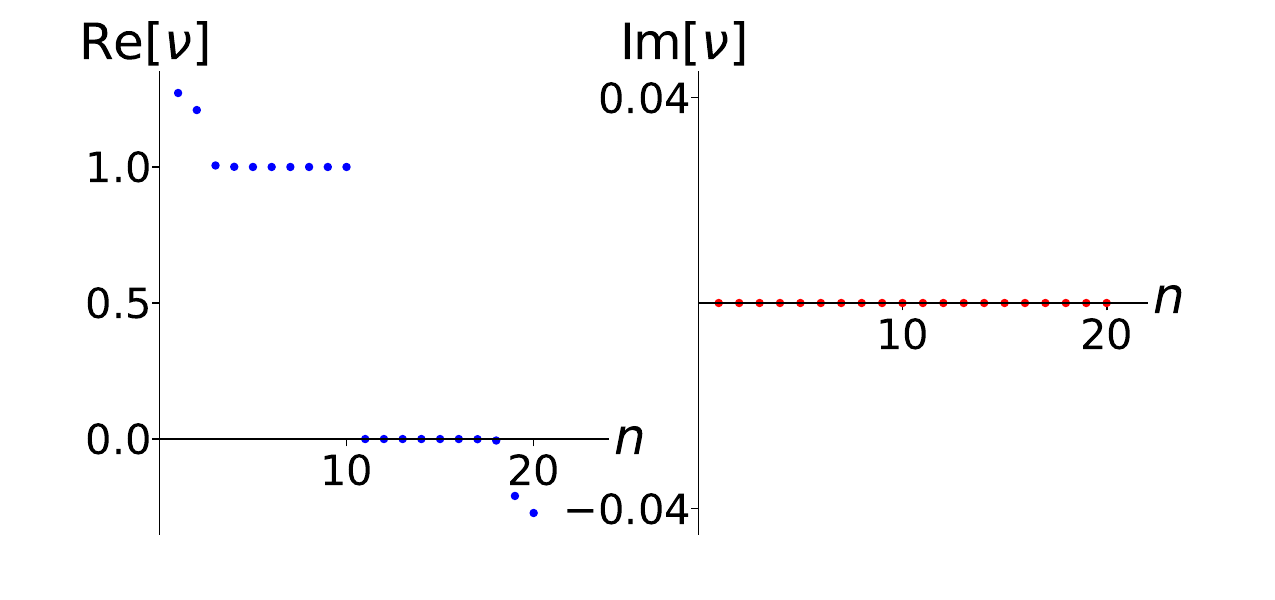}{4.1cm}{(a)}
	\end{minipage}
	\begin{minipage}[b]{0.48\linewidth}
		\panelH{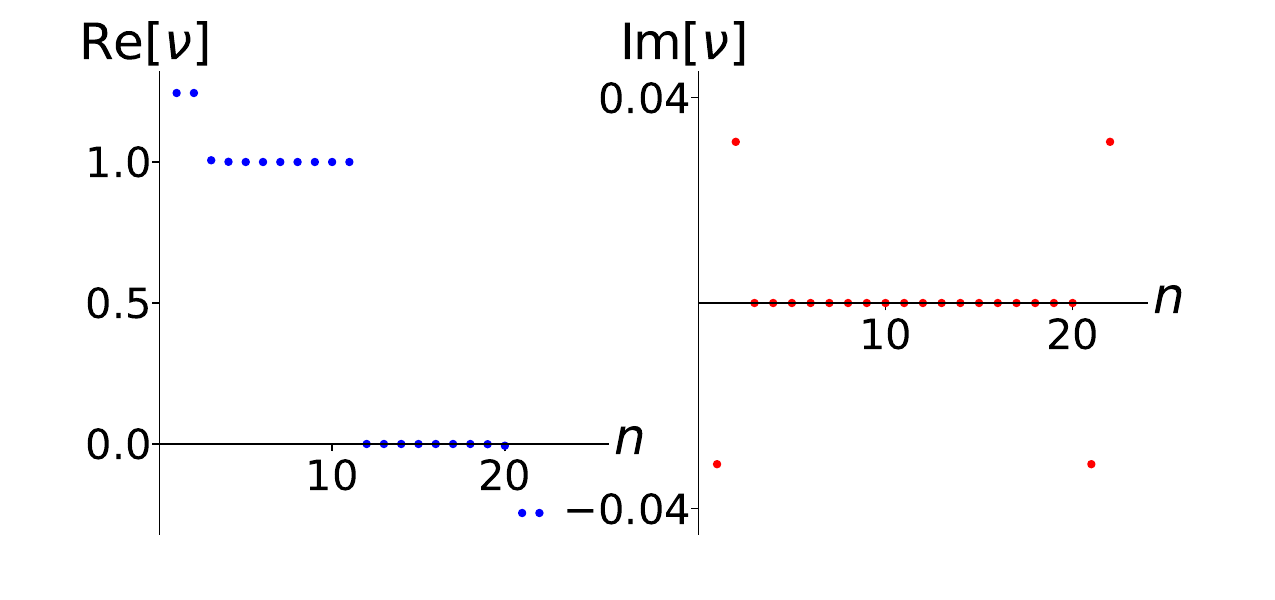}{4.1cm}{(b)}
	\end{minipage}\\
	\begin{minipage}[b]{0.298\linewidth}
		\panelH{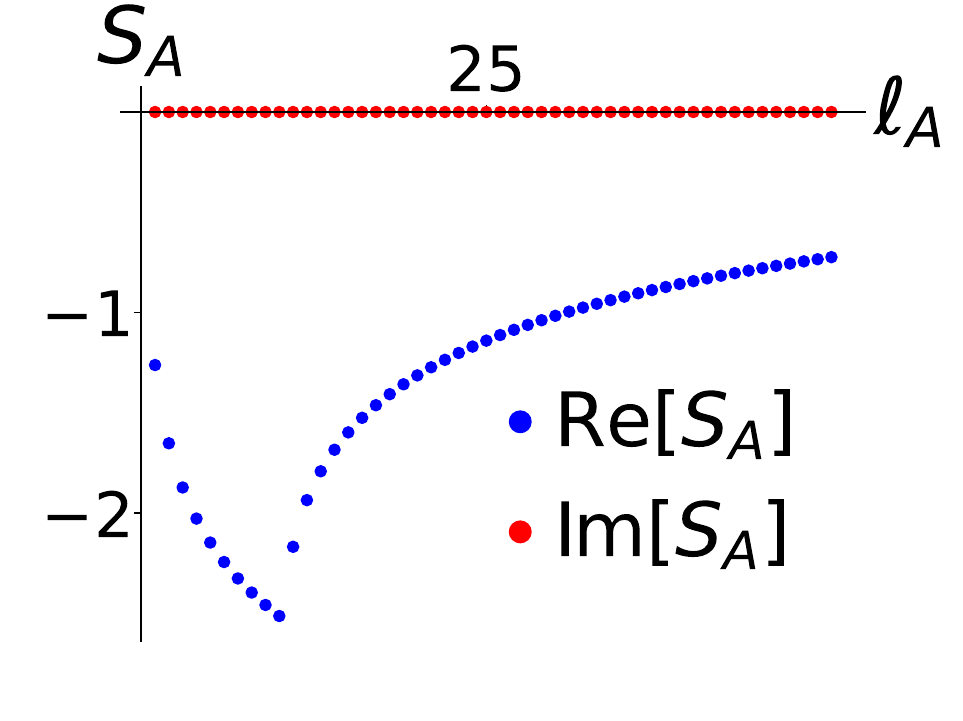}{4cm}{(c)}
	\end{minipage}
	\begin{minipage}[b]{0.298\linewidth}
		\panelH{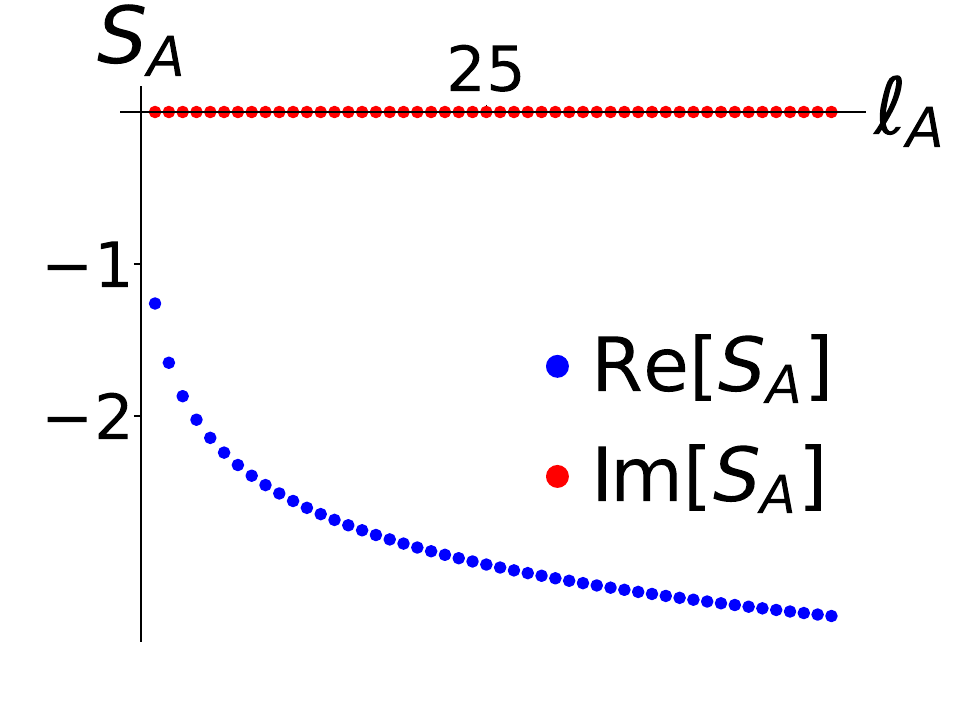}{4cm}{(d)}
	\end{minipage}
	\begin{minipage}[b]{0.33\linewidth}
		\panelH{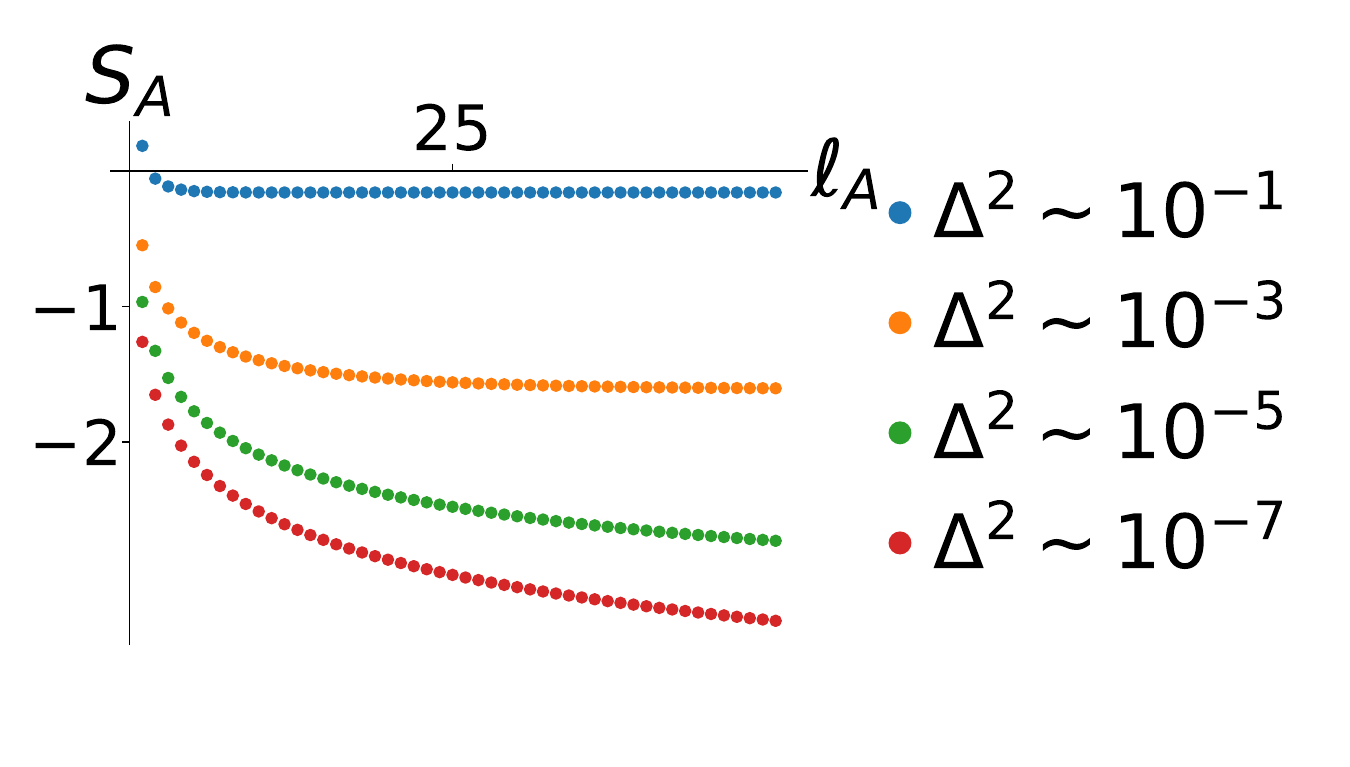}{4cm}{(e)}
	\end{minipage}
	\caption{Comparison of branch choices for entanglement in a small-gap non-Hermitian SSH chain
		($L=10^4$; $(v,w,u)=(2,1,1-10^{-7})$).
		(a–b) Entanglement spectrum $\{\nu\}$ around the onset: a complex quartet first appears at $\ell_A=11$ (absent at $\ell_A=10$).
		(c) Principal-branch \(S_A(\ell_A)\) shows an apparent "flip" at \(\ell_A=11\), coincident with the quartet onset.
		(d) Using a physically consistent branch removes the flip; for small gaps the scaling is consistent with \(c=-2\).
		(e) Correct-branch $S_A(\ell_A)$ for $u=1-10^{-7},\,1-10^{-5},\,1-10^{-3},\,1-10^{-1}$: the curves flatten smoothly with increasing gap $\Delta$ and approach the trivial entanglement (horizontal line).}
	\label{fig:gap_cal}
\end{figure}

The quartet contribution to the entanglement entropy is
\begin{equation}
S_{\rm quartet}
= -2\Big[\nu\ln\nu + (1-\nu^*)\ln(1-\nu^*)\Big] + \text{c.c.},
\label{eq:quartet-con}
\end{equation}
where $\nu=\mathcal{R}+i\,\mathcal{I}$ with $\mathcal{I}\in\mathbb{R_+}$. On the principal branch, write
\begin{equation}
\nu = re^{i\phi},\qquad 1-\nu^*=\rho e^{i\varphi},\qquad
r=|\nu|,\rho=|1-\nu|,
\phi=\arg(\nu),\varphi=\arg(1-\nu^*),\arg\in(-\pi,\pi].
\end{equation}
Substituting into \eqref{eq:quartet-con} gives
\begin{equation}
S_{\rm quartet}^{(\rm princ)}
=
-4\mathcal{R}\ln r
-4(1-\mathcal{R})\ln \rho
+4\mathcal{I}(\phi+\varphi),
\end{equation}
which is purely real on the principal branch. However, as noted above, numerics show an apparent "flip" in the entropy precisely when a complex quartet appears in the spectrum (see Fig.~\ref{fig:gap_cal} (a-c)), signaling a branch-choice artifact addressed below.

To resolve this branch-choice artifact, we adopt the same strategy used for the edge pair. The difference here is that a quartet contributes eight phases in total. Shifting the branches of two of these logarithms by $2\pi$ yields an expression analogous to Eq.~\eqref{eq:edge-branch}. In the quartet case the imaginary parts cancel exactly between conjugate contributions, and we obtain
\begin{equation}
S_{\rm quartet}^{(\rm branch)}
=
-4\mathcal{R}\ln r
-4(1-\mathcal{R})\ln \rho
+4\mathcal{I}\big(\phi+\varphi-\pi\big),
\end{equation}
which is purely real. With this branch assignment, the entropy recovers the smooth, gapped behavior expected in this regime (see Fig.~\ref{fig:gap_cal} (d-e)). 

\subsection{Symmetry constraints and the origin of conjugate pairs/quartets}
\label{subsec:symmetry-quartet}

Having shown that the entanglement entropy is well-defined whenever the correlation spectrum is symmetric, we now recall why this symmetry holds: microscopic symmetries enforce a pairing structure in the correlation spectrum~\cite{10.21468/SciPostPhys.7.5.069}. We write the Bloch Hamiltonian in the minimal $\mathcal{PT}$-symmetric form
\begin{equation}
H(k) = d_x(k)\sigma_x + d_y(k)\sigma_y + iu(k)\sigma_z,
\qquad d_x,d_y,u \in \mathbb{R}.
\end{equation}
There are two symmetries:

\emph{(i) time-reversal–like symmetry ($T_+$):}
\begin{equation}
H(k)^* = d_x\sigma_x - d_y\sigma_y - iu\sigma_z,
\end{equation}
and conjugation by $\sigma_x$ gives
\begin{equation}
\sigma_x H(k)^*  \sigma_x = d_x\sigma_x + d_y\sigma_y + iu\sigma_z = H(k).
\end{equation}

\emph{(ii) pseudo-Hermiticity ($PH_-$):}
\begin{equation}
H(k)^\dagger = d_x\sigma_x + d_y\sigma_y - iu\sigma_z,
\end{equation}
and conjugation by $\sigma_z$ yields
\begin{equation}
\sigma_z H(k)^\dagger  \sigma_z = -d_x\sigma_x - d_y\sigma_y - iu\sigma_z = -H(k).
\label{eq:ph-symmetry}
\end{equation}

These two Hamiltonian symmetries imply two constraints on the correlation spectrum, yielding the pairing structures used above.

\emph{Time–reversal–like ($T_+$) symmetry.}
Let $C_{ij} \equiv \bra{G_L} c_i^\dagger c_j\ket{G_R}$.
Using the standard identity for antiunitary maps,
\begin{equation}
\bra{G_L}\mathcal{O}\ket{G_R}^* = 
\bra{T_+G_L} T_+ \mathcal{O} T_+^{-1} \ket{T_+G_R}, \label{eq:non-un}
\end{equation}
together with $T_+ c T_+^{-1} = U_T c$ and $T_+ c^\dagger T_+^{-1} = c^\dagger U_T^\dagger$, we obtain
\begin{align}
(C^*)_{ij}
&= \bra{G_L} c_i^\dagger c_j \ket{G_R}^*
= \bra{T_+G_L} T_+ c_i^\dagger c_j T_+^{-1} \ket{T_+G_R} \nonumber\\
&= \bra{T_+G_L} (c^\dagger U_T^\dagger)_i  (U_T c)_j  \ket{T_+G_R}.
\end{align}
If the state is also invariant under time-reversal-like symmetry, we have
\begin{equation}
C = U_T  C^*  U_T^\dagger.
\end{equation}
The same relation holds for the subsystem matrix \(C_A\) whenever the restriction to \(A\) preserves the symmetry, i.e. the symmetry maps \(A\) to itself. (For the \(PT\) symmetry considered here, this requires choosing \(A\) to be symmetric about the inversion center.)
\begin{equation}
C_A = U_T  C_A^*  U_T^\dagger.
\end{equation}
Consequently, the spectrum is closed under complex conjugation:
\begin{equation}
\nu \leftrightarrow \nu^*.
\label{eq:ccsymmetry}
\end{equation}

\emph{Pseudo-Hermiticity ($PH_-$).}
Using Eq.~\eqref{eq:ph-symmetry}, $u_{\mathrm{ph}} H^\dagger u_{\mathrm{ph}}^\dagger=-H$ (with $u_{\mathrm{ph}}=\sigma_z$ in the Bloch basis), we now derive the corresponding constraint on the correlation matrix and the induced pairing of its eigenvalues.

Let $\{\,\ket{R_n}\ket{L_n}\}$ be the biorthogonal eigenvectors of $H$ and $H^\dagger$,
\begin{equation}
H\ket{R_n}=E_n\ket{R_n},\qquad H^\dagger\ket{L_n}=E_n^*\ket{L_n},
\qquad \braket{L_m}{R_n}=\delta_{mn},
\end{equation}
with completeness $\sum_n \ket{R_n}\bra{L_n}=\mathbf{I}$. From $u_{\mathrm{ph}} H^\dagger = - H u_{\mathrm{ph}}$ and $Hu_{\mathrm{ph}}^\dagger = - u_{\mathrm{ph}}^\dagger H^\dagger$, one finds the spectral pairing
\begin{equation}
E_{-n^*}=-E_n^*,\qquad
u_{\mathrm{ph}}\ket{R_n} = e^{i\alpha_n}\mathcal{N}_n \ket{L_{-n^*}},\qquad
u_{\mathrm{ph}}\ket{L_n} = e^{i\alpha_n}\mathcal{N}_n^{-1} \ket{R_{-n^*}},
\end{equation}
for some phases $e^{i\alpha_n}$ and nonzero normalizations $\mathcal{N}_n$ (fixed by gauge).
The correlation operator can be written as
\begin{equation}
C = \sum_n s_n\ket{R_n}\bra{L_n},
\end{equation}
with complex occupancies $s_n$; at biorthogonal half filling these satisfy $s_n^*+s_{-n^*}=1$ pairwise.

Conjugating $C^\dagger$ by $u_{\mathrm{ph}}$ and using the relations above,
\begin{equation}
u_{\mathrm{ph}}C^\dagger u_{\mathrm{ph}}^\dagger
=\sum_n s_n^*\ket{R_{-n^*}}\bra{L_{-n^*}}.
\end{equation}
Using $s_n^*+s_{-n^*}=1$ and completeness then gives the compact identity
\begin{equation}
u_{\mathrm{ph}}C^\dagger u_{\mathrm{ph}}^\dagger + C = \mathbf{I}.
\end{equation}
Because $u_{\mathrm{ph}}$ acts on-site, the same holds after restricting to the subsystem $A$:
\begin{equation}
u_{\mathrm{ph}}C_A^\dagger u_{\mathrm{ph}}^\dagger + C_A;= \mathbf{I}.
\end{equation}
Thus the spectrum is closed under the $PH_-$ mapping:
\begin{equation}  
\nu \leftrightarrow 1-\nu^*.
\end{equation}
Together with \eqref{eq:ccsymmetry} ($\nu\leftrightarrow\nu^*$), this ensures that every complex eigenvalue groups into the required pairs or fourfold sets, underpinning our branch-cut prescription.

\subsection{Comparison with the absolute-value prescription}

An alternative prescription from prior work~\cite{Tu2022SciPost} reproduces the central charge at the nH SSH topological QCP by replacing the multivalued complex logarithm in the entropy with $\ln {\rm abs}{\,\cdot\,}$,
\begin{equation}
S_A^{(\mathrm{abs})}
= -\sum_n \big[\nu_n \ln|\nu_n| + (1-\nu_n)\ln|1-\nu_n|\big],
\end{equation}
i.e. effectively $S=-\mathrm{Tr}\rho\ln\rho \mapsto -\mathrm{Tr}\rho\ln|\rho|$ at the correlation-spectrum level. For the edge pair $\nu_\pm=\tfrac{1}{2}\pm i\mathcal{I}$ this gives
\begin{equation}
S_{\rm edge}^{(\mathrm{abs})}=-2\ln r,\qquad r=\sqrt{\tfrac14+\mathcal{I}^2},
\end{equation}
whereas our branch-cut prescription yields
\begin{equation}
S_{\rm edge}^{(\mathrm{branch})}
= -2\ln r + (4\phi-2\pi)\mathcal{I} - i\pi,
\qquad \phi=\arctan(2\mathcal{I}).
\end{equation}
Thus the absolute-value method removes the $4\phi\mathcal{I}$ term by construction. To compare the two methods, we therefore examine the real part of $S_A$, the difference is
\begin{equation}
\Delta\Re S_{\rm edge}
\equiv
\Re S_{\rm edge}^{(\mathrm{branch})}-S_{\rm edge}^{(\mathrm{abs})}
=(4\phi-2\pi)\mathcal{I}.
\end{equation}

\begin{figure}[htbp]
	\centering
	\begin{minipage}[b]{0.43\linewidth}
		\panelH{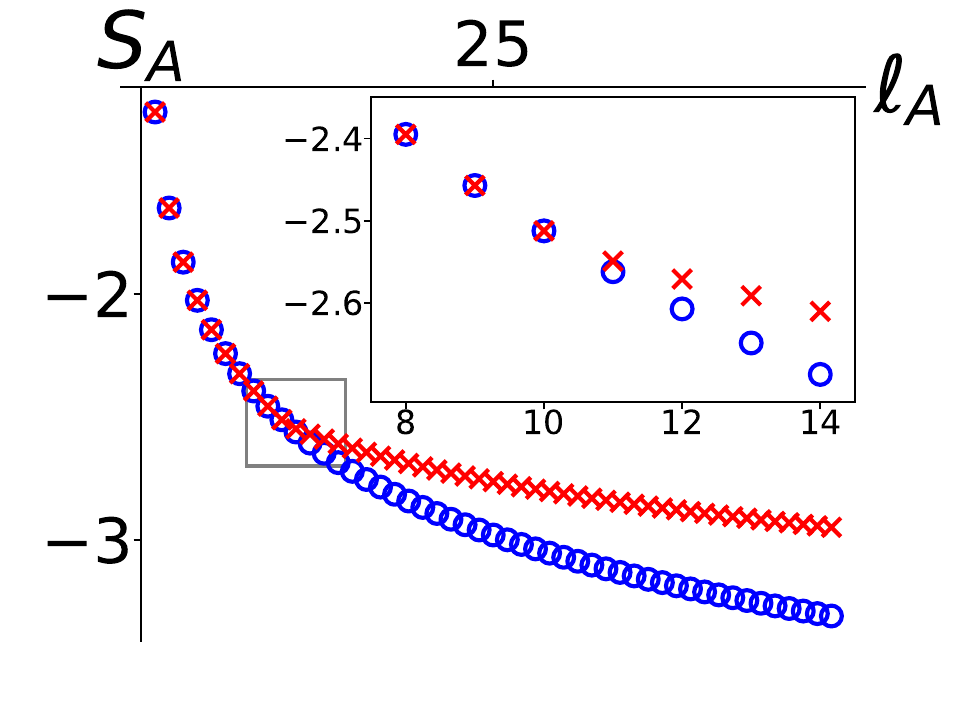}{4.5cm}{(a)}
	\end{minipage}
	\begin{minipage}[b]{0.54\linewidth}
		\panelH{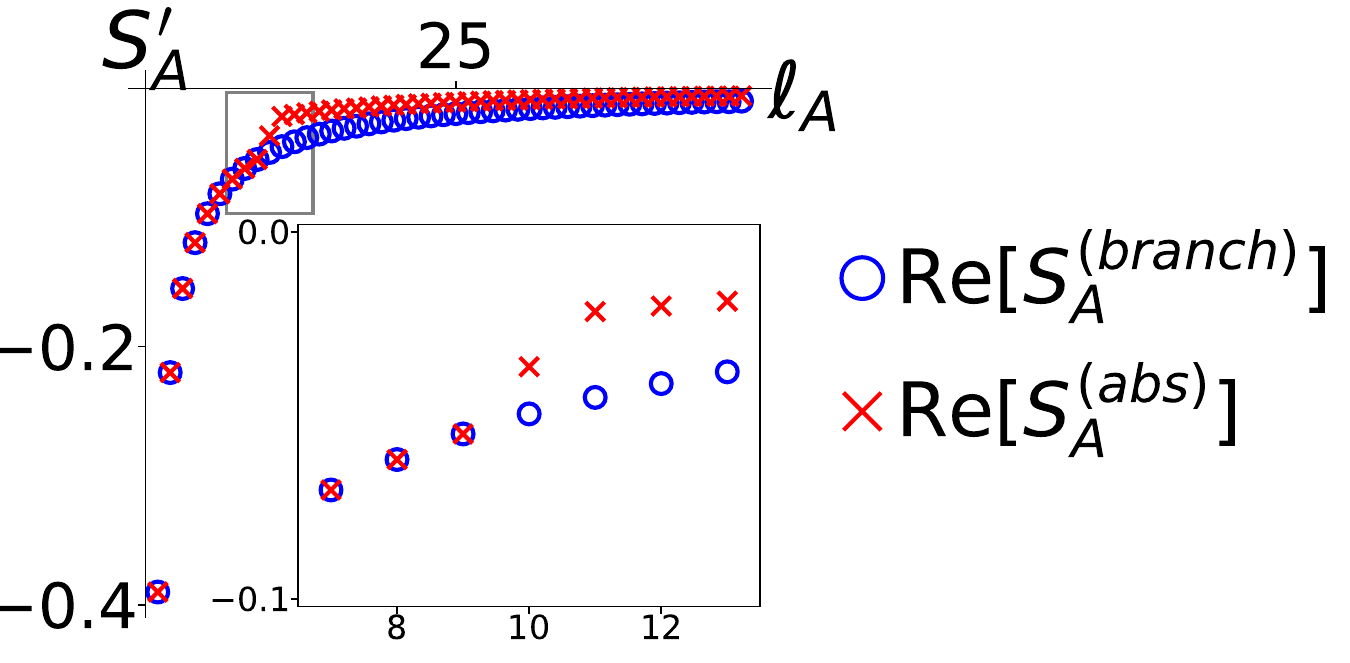}{4.5cm}{(b)}
	\end{minipage}\\
    \begin{minipage}[b]{0.32\linewidth}
    	\panelH{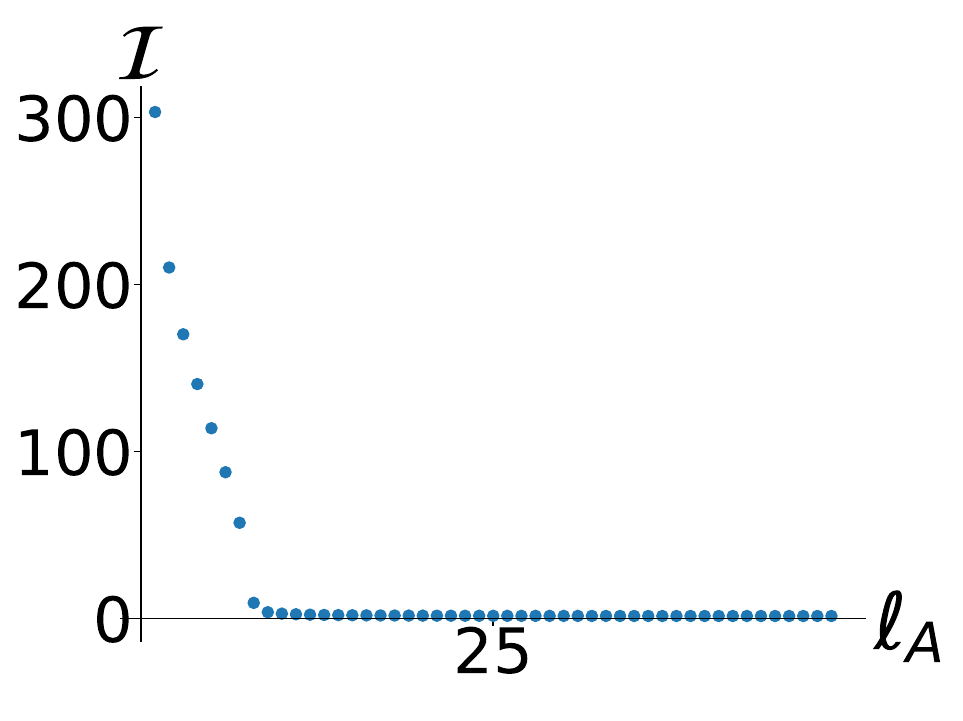}{4.3cm}{(c)}
    \end{minipage}
	\begin{minipage}[b]{0.33\linewidth}
		\panelH{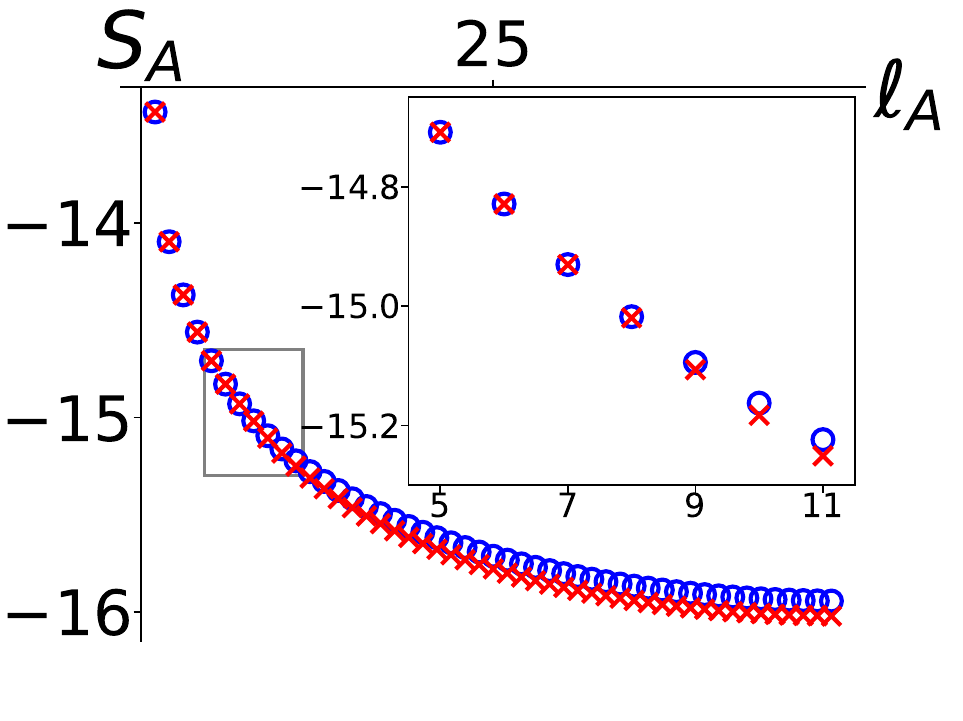}{4.3cm}{(d)}
	\end{minipage}
	\begin{minipage}[b]{0.33\linewidth}
		\panelH{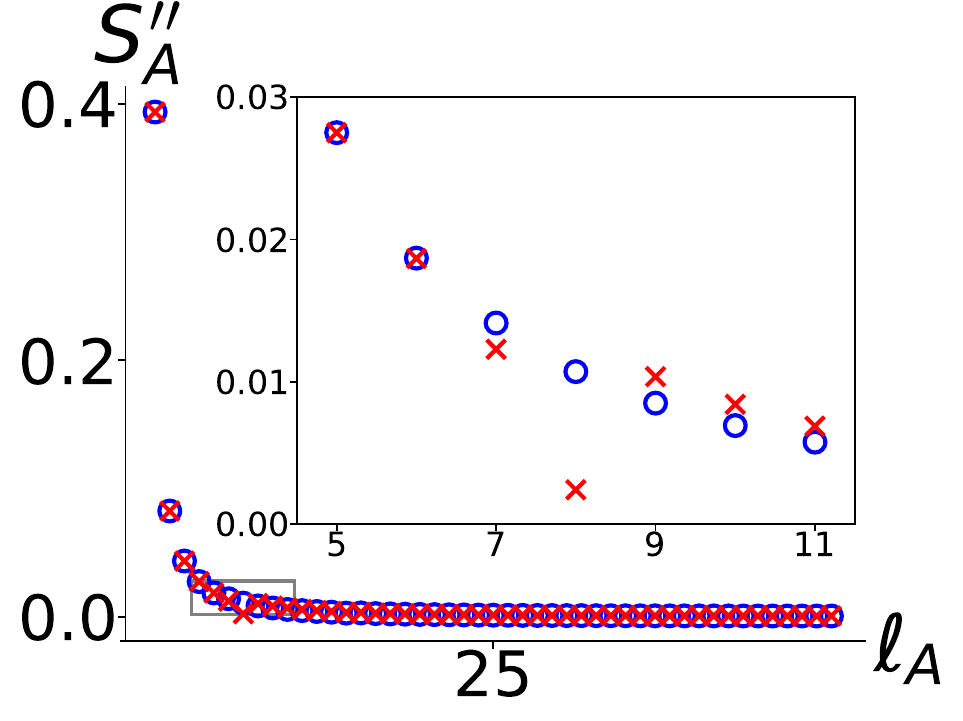}{4.3cm}{(e)}
	\end{minipage}
	\caption{Comparison of entanglement–entropy prescriptions: branch-cut vs absolute value.
		Red $\times$ denotes $\mathrm{Re}[S_A^{(\mathrm{abs})}]$, and blue $\circ$ denotes $\mathrm{Re}[S_A^{(\mathrm{branch})}]$. Small-gap nH SSH, $(L=10000; (v,w,u)=(2,1,1-10^{-7}))$.
		A complex quartet first appears at $\ell_A=11$ (see Fig. \ref{fig:gap_cal}(a-b)).
		(a) $\mathrm{Re}[S_A]$: the absolute-value prescription
		$\mathrm{Re}[S_A^{(\mathrm{abs})}]$ shows a kink at $\ell_A=11$, whereas the branch-cut result $\mathrm{Re}[S_A^{(\mathrm{branch})}]$ remains smooth.
		(b) Slope $S_A'$: a clear jump occurs only for the absolute-value curve. Higher-$\alpha$-nH SSH $(L=100; \alpha=2; (v,w,u)=(3,1,2-10^{-15}))$.
		(c) Edge-mode imaginary part $\mathcal{I}(\ell_A)$ drops rapidly and becomes $\mathcal{O} (1)$ for $\ell_A \geq 8$.
		(d) $\mathrm{Re}[S_A]$: once $\mathcal{I}$ is $\mathcal{O} (1)$, the two prescriptions diverge smoothly; for visual alignment we subtract the constant offset from the absolute-value curve.
		(e) Second derivative $S_A''$: a spike at $\ell_A=8$ is present for the absolute-value prescription, whereas the branch-cut result remains smooth.	
	}
	\label{fig:abs_comp}
\end{figure} 
At the topological QCP we have $\mathcal I\to\infty$, hence $\phi\to\pi/2$. Thus $\Delta\Re S_{\rm edge}$ tends to a finite constant $-2$; up to this constant offset, the two prescriptions agree on the real scaling. While the two prescriptions coincide for the edge mode at the topological QCP of the nH SSH model, the situation is different in more general settings. Whenever a complex mode has a finite imaginary part, $\mathcal{I}=\mathcal{O}(1)$ (for example, in gapped modes), the difference is nonzero and the two prescriptions no longer agree. For gapped modes with $\nu=\mathcal{R}+i\mathcal{I}$,
\begin{equation}
S_{\rm quartet}^{(\mathrm{abs})}
= -4\mathcal{R}\ln r - 4(1-\mathcal{R})\ln \rho,
\qquad
S_{\rm quartet}^{(\mathrm{branch})}
= S_{\rm quartet}^{(\mathrm{abs})} + 4\mathcal{I}(\phi+\varphi-\pi),
\end{equation}
with $r=|\nu|$, $\rho=|1-\nu|$, $\phi=\arg\nu$, and $\varphi=\arg(1-\nu^*)$. The discrepancy stems from this quartet contribution, so a clear difference appears as soon as a quartet first enters the entanglement spectrum. At that point the absolute-value prescription typically shows a kink, whereas the branch-cut prescription remains smooth (see Fig.~\ref{fig:abs_comp}(a-b)).

As an illustration in higher–$\alpha$ models, edge modes may emerge without mass inversion (see SM Sec.~\ref{sm5}). These modes typically do not satisfy $\mathcal{I}\gg1$ but instead saturate to a finite value. In the $\alpha=2$ nH–SSH chain at its lower–winding QCP, the edge–mode imaginary part $\mathcal I(\ell_A)$ starts large and then drops rapidly, becoming $\mathcal O(1)$ at $\ell_A=8$ (Fig.~\ref{fig:abs_comp}(c)). In this regime the two prescriptions diverge smoothly, producing a continuous offset in $\mathrm{Re}[S_A]$ (Fig.~\ref{fig:abs_comp}(d)). This appears as a spike in the second derivative $S_A''$ at $\ell_A=8$ (Fig.~\ref{fig:abs_comp}(e)).

In summary, both prescriptions reproduce the real scaling at the topological QCP of the ordinary nH–SSH chain (up to an irrelevant constant offset), but the branch-cut prescription preserves the original definition $S=-\mathrm{Tr}\,\rho\ln\rho$ and is less sensitive to small gaps. Moreover, in higher-$\alpha$ models with non–mass-inversion edge modes—where the edge-mode imaginary part saturates to $\mathcal O(1)$—the absolute-value prescription introduces non-smooth behavior in the second derivative around this crossover. By contrast, the branch-cut result remains smooth. Taken together, the branch-cut prescription is the more convincing and physically grounded method for computing the nH entanglement entropy.

\subsection{Interaction effect-the staggered XXZ model}
We consider the nearest-neighbor Hubbard interaction in the nH SSH model. The interaction term is given by
\begin{align}
H_U= \sum_{i} U (n_{A,i} n_{B,i} + n_{B i} n_{A,i+1}),
\end{align}
where $n_{A/B,i} = c^\dagger_{A/B,i} c_{A/B,i}$ with $c^{(\dagger)}_{A/B,i}$ being the fermion operator at site $i$ on sublattice $A/B$.
One can perform the Jordan-Wigner on the fermionic model to map it to a spin model. The corresponding spin model is 
\begin{align}
H=\sum_i v (X_{A,i}X_{B,i}+Y_{A,i}Y_{B,i})- w (X_{B,i}X_{A,i+1}+Y_{B,i}Y_{A,i+1})+i u (Z_{A,i}-Z_{B,i}) + U (Z_{A,i}Z_{B,i}+Z_{B,i}Z_{A,i+1}),
\end{align}
which is the XXZ model with staggered interactions between sublattices $A$ and $B$.

We consider the many-body ground state of the staggered XXZ model and analyze eigenstates and  eigenvalues of the biothogonal reduced density matrix.  
In the non-interacting limit, the reduced density matrix can be expressed as the tensor product form as
\begin{align}
\rho_A = \otimes_i \begin{pmatrix}
\nu_i & 0 \\
0 & 1-\nu_i 
\end{pmatrix}=
\begin{pmatrix}
\xi_1 & 0 & \cdots &0 \\
0 &  \xi_2 & \cdots & 0 \\   \vdots & & \ddots   \\ 0  
 & & &\xi_N       
\end{pmatrix}.
\end{align}

That is, we have $ {\rm Tr }[\rho_A]=
\sum_j\xi_j = \prod_i (\nu_j + (1-\nu_i)) =1$. 

\begin{figure}[htbp]
\includegraphics[width=0.7\textwidth]{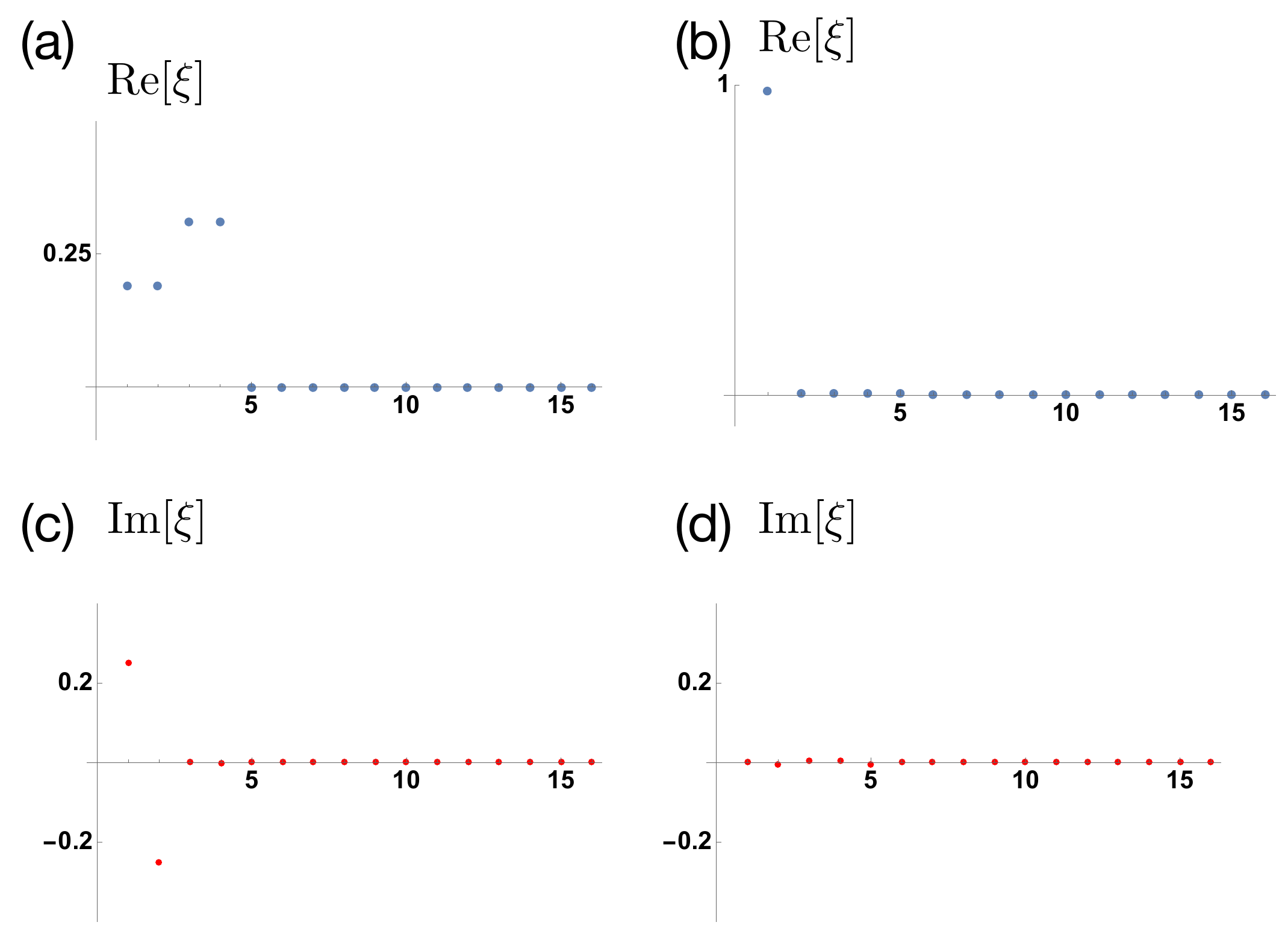}
\caption{Upper panel: the real part of the eigenvalues of the reduced density matrix of the staggered XXZ model.
Lower panel: the imaginary part of the eigenvalues of the reduced density matrix of the staggered XXZ model.(a)(c) The parameters (v,w,u,U)=(1,3.2,2.2,0.5).(b)(d) The parameters (v,w,u,U)=(3.2,1,2.2,0.5). The total number of sites is ten and the subsystem $A$ contains four sites which have $16$ eigenvalues.} \label{fig:nHXXZ}
\end{figure}

We can associate the entangling boundary modes in the correlation matrix with those in the reduced density matrix via Eq.~(\ref{Eq:eco}).
\begin{align}
\rho_A = 
\begin{pmatrix}
\frac{1}{4}+ \mathcal{I}_1^2 & 0 &0 &0  \\
0 & \frac{1}{4}- \mathcal{I}_1^2 +i \mathcal{I}_1 &0&0 \\
0&0& \frac{1}{4}- \mathcal{I}_1^2 -i \mathcal{I}_1 &0 \\
0&0&0& & \frac{1}{4}+ \mathcal{I}_1^2
\end{pmatrix}
\otimes_{j\neq 1,2} \begin{pmatrix}
\nu_j & 0 \\
0 & 1-\nu_j 
\end{pmatrix},
\label{Eq:mm}
\end{align}
where $\nu_{1} = \frac{1}{2} + i \mathcal{I}_1 $ and  $\nu_{2} = \frac{1}{2} - i \mathcal{I}_1 $ are two entangling boundary modes in the correlation matrix. This leads to four entangling modes in the reduced density matrix, where the real part of the corresponding eigenvalues are ${\rm Re} [\xi_j] \sim 0.25 \pm \mathcal{I}_1^2 $.These modes form complex conjugation pairs in the presence of the nonvanishing $U$
as shown in Figs.~\ref{fig:nHXXZ}(a)(c).   However, in the trivial gapped region, there are no entangling boundary modes as shown in Figs.~\ref{fig:nHXXZ}(b)(d).

For the many-body formula of the branch choice in logarithmic function of the reduced density matrix, we can trace it back to the branch  choice in the free-fermion case [Eq.~(\ref{eq:edge-branch})] and implement it in the matrix form in Eq.~(\ref{Eq:mm}). In the free fermion limit, it is 
\begin{align}
\ln \rho_A =\ln \left(  
\begin{pmatrix}
(\frac{1}{4}+ \mathcal{I}_1^2) e^{i 2 \pi} & 0 &0 &0  \\
0 & (\frac{1}{4}- \mathcal{I}_1^2 +i \mathcal{I}_1) e^{i 2 \pi} &0&0 \\
0&0& \frac{1}{4}- \mathcal{I}_1^2 -i \mathcal{I}_1 &0 \\
0&0&0& & \frac{1}{4}+ \mathcal{I}_1^2
\end{pmatrix}
\otimes_{j\neq 1,2} \begin{pmatrix}
\nu_j & 0 \\
0 & 1-\nu_j 
\end{pmatrix} \right).
\label{Eq:mm}
\end{align}
The above form indicates that in the presence of these four entangling boundary modes, the spectrum of the reduced density matrix must come in conjugation pairs, 
${\rm spec} (\rho_A) = \{ \xi_1, \xi^*_1, \cdots, \xi_n, \xi^*_n \}$.
The branch choice for computing the entanglement entropy is that the complex conjugation partners take an additional $2 \pi$ phase. That is,
\begin{align}
 S_A = - \sum_i \xi_i \ln \xi_i   - \xi_i^* \ln (\xi^*_i e^{i 2 \pi})=-\sum_i \xi_i \ln \xi_i   - \xi_i^* \ln \xi^*_i - \xi_i^* (2 \pi i).
\end{align}
The imaginary part of the entanglement entropy is
\begin{align}
{\rm Im} ( S_A)  = - \sum_i \xi_i^* (2 \pi i)= - i \pi.
\end{align}
Here we use $\sum_i \xi_i +\xi_i^* =1$ and ${\rm Re} (\sum_i \xi_i)={\rm Re} (\sum_i \xi^*_i)= \frac{1}{2} $.

Now we can generalize this to  the many-body formula with the topological number  $\Omega$. The number of the entangling boundary modes in the reduced density is $4^\Omega$. Based on observations of the structure of the entangling boundary states in the correlation matrix, we conjecture that the spectrum of the reduced matrix has $2^{\Omega-1}$ group of conjugation pairs denoted by ${\rm Spec} (\rho_A) = \{ (\xi_{1,\alpha},\xi_{2,\alpha}=\xi^*_{1,\alpha}),(\xi_{3,\alpha},\xi_{4,\alpha}=\xi^*_{3,\alpha}),\cdots,(\xi_{2^\Omega-1,\alpha},\xi_{2^\Omega,\alpha}=\xi^*_{2^\Omega-1,\alpha})\}$.
These group of conjugation pairs satisfy
\begin{align}
{\rm Re} (\sum_\alpha \xi_{j,\alpha} = \frac{1}{2^\Omega}), \quad j=1,\cdots, 2^\Omega
\end{align}

It leads to ${\rm Tr}(\rho_A)=\sum_{j,\alpha} \xi_{j,\alpha}= 1$. 
The branch choice for computing the entanglement entropy
is
\begin{align}
S_A =& -\sum_\alpha [(\xi_{1,\alpha} \ln \xi_{1,\alpha} + \xi_{2,\alpha} \ln (\xi_{2,\alpha}e^{i 2 \pi \Omega}) )+(\xi_{3,\alpha} \ln (\xi_{3,\alpha}e^{i 2 \pi}) + \xi_{4,\alpha} \ln (\xi_{4,\alpha}e^{i 2 \pi (\Omega-1)}) ) \notag\\
& +\cdots+(\xi_{2^\Omega-1,\alpha} \ln (\xi_{2^\Omega-1,\alpha}e^{i 2 \pi \frac{\Omega}{2}}) + \xi_{2^\Omega,\alpha} \ln (\xi_{2^\Omega,\alpha}e^{i 2 \pi \frac{\Omega}{2}}) )], \quad \Omega \in 2\mathbb{Z}.
\end{align}
And 
\begin{align}
S_A =& -\sum_\alpha [(\xi_{1,\alpha} \ln \xi_{1,\alpha} + \xi_{2,\alpha} \ln (\xi_{2,\alpha}e^{i 2 \pi \Omega}) )+(\xi_{3,\alpha} \ln (\xi_{3,\alpha}e^{i 2 \pi}) + \xi_{4,\alpha} \ln (\xi_{4,\alpha}e^{i 2 \pi (\Omega-1)}) ) \notag\\
& +\cdots+(\xi_{2^\Omega-1,\alpha} \ln (\xi_{2^\Omega-1,\alpha}e^{i 2 \pi \frac{\Omega-1}{2}}) + \xi_{2^\Omega,\alpha} \ln (\xi_{2^\Omega,\alpha}e^{i 2 \pi \frac{\Omega+1}{2}}) )], \quad \Omega \in 2\mathbb{Z}+1.
\end{align}

We can take a closer look at the imaginary part of the entanglement entropy, which is 
\begin{align}
{\rm Im} (S_A) &= - \frac{1}{2^\Omega} 2\pi (\Omega + \Omega(\Omega-1)+ \frac{\Omega(\Omega-1)(\Omega-2)}{2}+ \frac{\Omega(\Omega-1)(\Omega-2)(\Omega-3)}{3!}+\cdots + \frac{\Omega(\Omega-1)\cdots}{(\Omega-1)!} ) \notag\\
& = - \pi 2^{1-\Omega }\Omega\sum_{k=0}^{\Omega-1} \binom{\Omega-1}{k}= - \pi 2^{1-\Omega }\Omega 2^{\Omega-1} = - \Omega \pi.
\end{align}

The above result agree with the branch choice for computing the entanglement entropy using the correlation matrix method in free-fermion cases. Here, we conjecture that in the presence of entangling boundary modes in the reduced density matrix with interactions, the spectrum of the reduced density matrix forms $2^{\Omega-1}$ group of conjugation pairs, which originate from its entangling boundary modes.

\section{Entanglement–edge correspondence}
\label{sm5}
In this section we record an entanglement--edge correspondence that appears in the $\mathcal{PT}$-symmetric non-Hermitian free-fermion model investigated in this work. We find that boundary modes in the physical spectrum under OBC have direct counterparts in the entanglement spectrum under PBC.

\begin{figure}[t]
\centering 
\begin{minipage}[b]{0.46\linewidth} 
\panel{ES_norm-eps-converted-to.pdf}{\linewidth}{(a)} 
\end{minipage}
\begin{minipage}[b]{0.46\linewidth} 
\panel{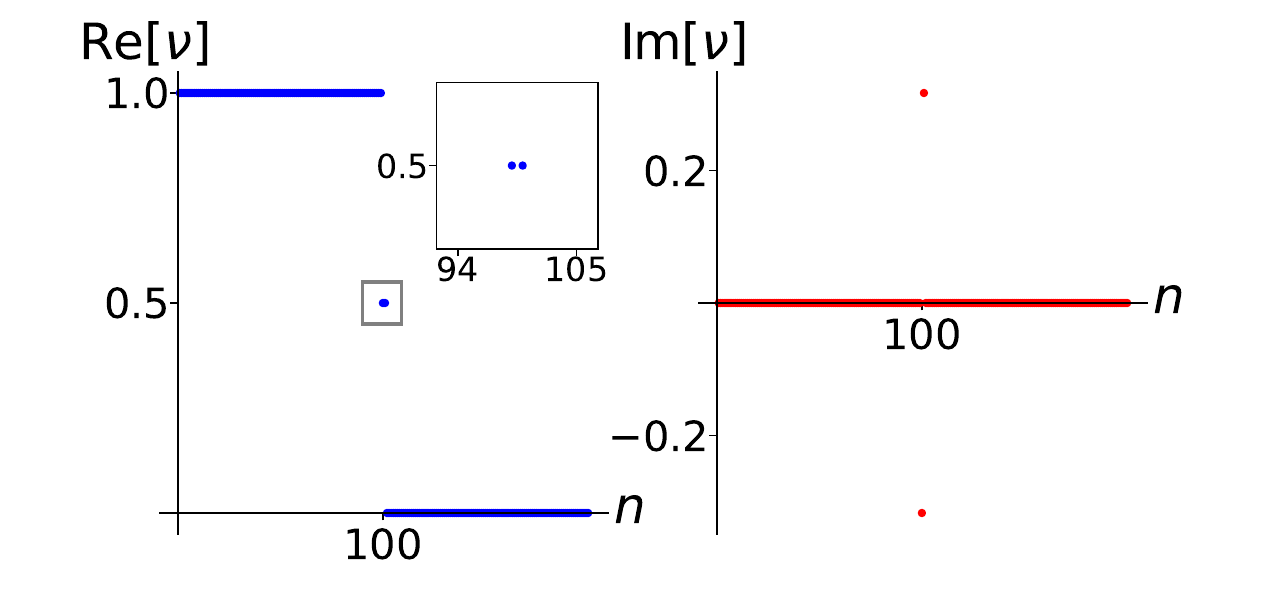}{\linewidth}{(b)} 
\end{minipage}\\
\begin{minipage}[b]{0.46\linewidth} 
\panel{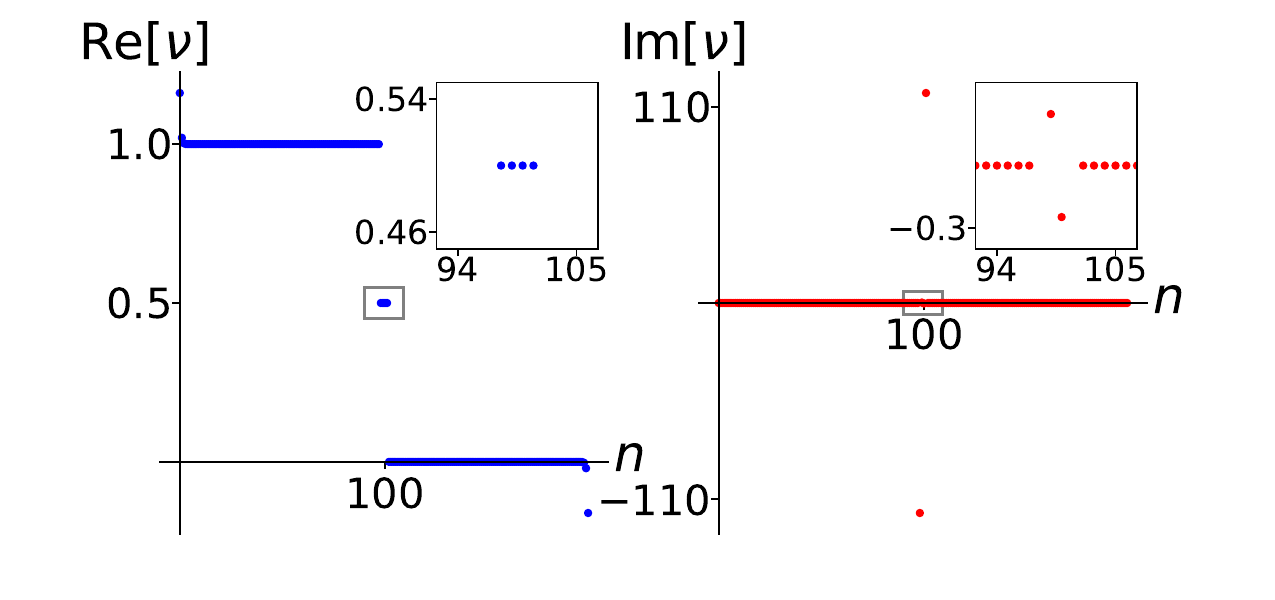}{\linewidth}{(c)} 
\end{minipage} 
\begin{minipage}[b]{0.46\linewidth} 
\panel{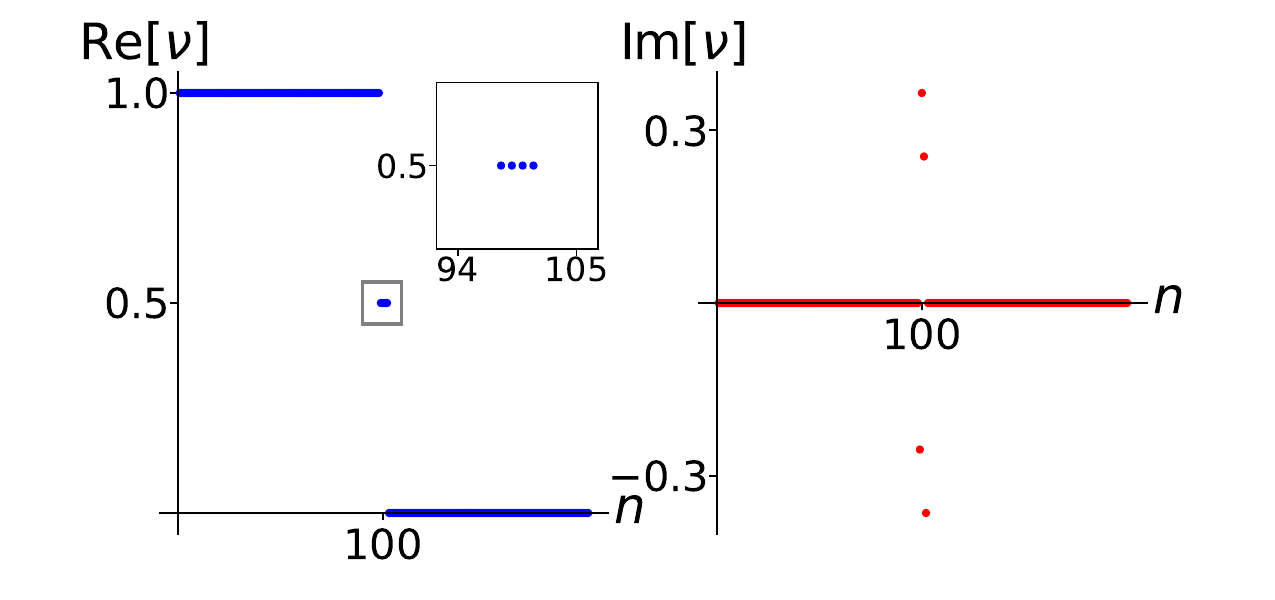}{\linewidth}{(d)} 
\end{minipage} 
\caption{
Entanglement spectrum of the $\alpha$-extended nH SSH model at topological criticality and in the corresponding gapped phases.
In each subfigure, the left (right) panel shows the real (imaginary) part, $\mathrm{Re}[\nu]$ ($\mathrm{Im}[\nu]$), of the correlation-spectrum eigenvalues.
(a) $\alpha=1$ at the topological QCP
(b) $\alpha=1$ in the gapped topological phase
(c) $\alpha=2$ ($\Omega=2$) at the topological QCP
(d) $\alpha=2$ ($\Omega=2$) in the gapped topological phase
Across both $\alpha=1,2$ and both regimes (critical and gapped), the number of complex-conjugate pairs in the correlation spectrum matches the number of $\mathcal{PT}$-protected edge modes.
} \label{fig:edge-cor}
\end{figure}

Let us start from the physical spectrum. As stated in the main text, we observe a bulk--edge correspondence in this model: for \(\mathcal{PT}\)-symmetric topological phase with winding number \(\Omega\), the OBC spectrum contains \(\Omega\) pairs of edge-localized modes pinned at purely imaginary energies,
\begin{equation}
E_{\rm edge}=\pm i u .
\end{equation}
We next examine the entanglement spectrum under PBC. We compute the correlation matrix for a sufficiently long contiguous interval \(A\) and diagonalize the restricted correlator \(C_A\). In the \(\mathcal{PT}\)-symmetric topological phase with winding number \(\Omega\), the resulting correlation spectrum exhibits \(\Omega\) characteristic complex-conjugate pairs of eigenvalues of the form
\begin{equation}
\nu_a=\frac{1}{2}\pm i\,\mathcal{I}_a,\qquad a=1,\dots,\Omega,
\label{Eq:eco}
\end{equation}
which are precisely the entanglement edge-mode pairs discussed in Sec.~\ref{subsec:edge-branch}. Such a correspondence appears to be a generic feature of \(\mathcal{PT}\)-symmetric non-Hermitian free-fermion chains. This correspondence is illustrated in Fig.~\ref{fig:edge-cor}: panels (a,b) show a representative \(\Omega=1\) example, while panels (c,d) show the \(\Omega=2\) case. Moreover, the \(\mathcal{PT}\)-preserving disorder results presented in the main text indicate that the correspondence does not rely on translational invariance.

It is also useful to translate these correlation-spectrum pairs into entanglement energies. For free fermions the single-particle entanglement energies are related to correlation eigenvalues by
\begin{equation}
\nu=\frac{1}{1+e^{\epsilon}}
\quad \Longleftrightarrow\quad
\epsilon=\ln\frac{1-\nu}{\nu}.
\end{equation}
Plugging in $\nu=\tfrac12\pm i\mathcal{I}$ gives
\begin{equation}
\epsilon=\ln\frac{1-\nu}{\nu}
=\pm\,2i\,\arctan(2\mathcal{I}),
\end{equation}
so these special pairs correspond to entanglement levels whose energies are purely imaginary. In this way, even though both the physical spectrum (under OBC) and the entanglement spectrum (under PBC) are complex in the non-Hermitian setting, the modes participating in this correspondence are ``zero-real-part'' modes on both sides: the OBC boundary modes sit at $\mathrm{Re}\,E=0$, and the corresponding entanglement levels sit at $\mathrm{Re}\,\epsilon=0$. This structure is reminiscent of the Li--Haldane entanglement--edge correspondence in Hermitian SPT chains.

\section{Interpretation of quantized imaginary entanglement entropy}
\label{sm6}

\subsection{Quantized imaginary entropy as a topological index}

As discussed in the main text, requiring the entanglement entropy to exhibit the expected conformal scaling---consistent with a non-unitary CFT description---along the $\mathcal{PT}$-symmetric topological critical lines of the $\alpha$-extended non-Hermitian SSH family effectively fixes the branch choice of the logarithms. Adopting this prescription, detailed in Sec.~\ref{subsec:edge-branch}, the entanglement entropy under PBC takes the $\Omega$-resolved form
\begin{equation}
S_A(\ell_A)\;=\;-\frac{2}{3}\ln \ell_A\;-\; i\,\Omega\,\pi\;+\;\mathrm{const.},
\label{eq:SM_EE_topo_const}
\end{equation}
where $\ell_A$ is the subsystem length and the constant includes nonuniversal UV contributions. Notably, the imaginary part is a quantized constant $-i\Omega\pi$ determined solely by the winding number $\Omega$.

It is natural to extend the same (conformally consistent) branch prescription into the adjacent $\mathcal{PT}$-symmetric SPT phase by continuity. With this extension, $\operatorname{Im}S_A$ remains locked to the same constant throughout each fixed-$\Omega$ sector, so that $\operatorname{Im}S_A$ can be viewed as a topological index reflecting the winding number $\Omega$ across the $\mathcal{PT}$-symmetric sector, thereby encoding the associated long-range physics.

\subsection{Boundary-entropy interpretation}

A natural framework for universal constant terms in 1D criticality is the Affleck--Ludwig boundary entropy (the $g$-function)~\cite{PhysRevLett.67.161}. In boundary CFT, a conformal boundary condition is represented by a boundary state $|B\rangle$, and the associated boundary amplitude is the vacuum overlap $g=\langle 0|B\rangle$. For an interval of length $\ell$ in a critical system, the entanglement entropy then takes the BCFT form \cite{Ohmori_2015}
\begin{equation}
S(\ell)=\frac{c}{3}\ln(\ell/\epsilon)+\ln g_a+\ln g_b,
\label{eq:SM_BCFT_EE}
\end{equation}
where $\epsilon$ is a short-distance cutoff, and $a,b$ label the conformal boundary conditions at the two endpoints.

When we consider entanglement under PBC, the ``boundaries'' in Eq.~\eqref{eq:SM_BCFT_EE} are created by the bipartition (the entanglement cuts). Since the two endpoints are generated by the same cut construction, it is natural that they realize the same effective boundary condition, and hence $g_a=g_b\equiv g$. A crucial point is that the conformal boundary condition associated with an entanglement cut need not be trivial. In symmetry-enriched quantum criticality, bulk topological structure can obstruct a \emph{na\"{\i}ve} symmetry-preserving boundary from being trivial, so that a nontrivial conformal boundary state may be naturally realized~\cite{Yu2022PRL}.

With this framework in mind, we return to our observation: the quantized imaginary term $-i\Omega\pi$ is (i) independent of $\ell_A$, (ii) robust under UV deformations, and (iii) takes distinct quantized values for different $\Omega$. These are precisely the qualitative features expected of an IR fixed-point boundary-entropy contribution, i.e.\ the $2\ln g$ term in Eq.~\eqref{eq:SM_BCFT_EE}. Taking the topological QCP of the nH SSH model as an example ($\Omega=1$), although the real part of the boundary entropy is hard to extract due to nonuniversal UV contributions, the imaginary part is fixed and yields
\begin{equation}
2\ln g \;=\; \text{(some real part)} \;-\; i\pi
\quad \Rightarrow \quad
g\in i\mathbb{R}.
\end{equation}
This implies that the corresponding boundary amplitude is imaginary. Following the original interpretation of the Affleck--Ludwig $g$-function as a boundary degeneracy \cite{PhysRevLett.67.161}, one may view this result as indicating an \emph{imaginary} boundary degeneracy associated with the non-Hermitian boundary state. More generally, if multiple edge modes are present, their contributions combine multiplicatively in the boundary amplitude, consistent with the observed $\Omega$-dependent quantization of the imaginary constant $-i\Omega\pi$.

Finally, an imaginary boundary amplitude is not in conflict with the boundary-entropy framework once unitarity is relaxed: the standard positivity properties of $g$ in conventional BCFT rely on unitarity/reflection-positivity assumptions~\cite{Friedan_2006}, whereas non-unitary or non-Hermitian critical points do not generically obey those constraints.

\subsection{R\'enyi entropy as supportive evidence}
\label{subsec:SM_imagEE_C}

The boundary-entropy contribution has a useful diagnostic feature: in the BCFT expressions for the R\'enyi entropies, the endpoint contribution enters as an \emph{additive constant} and, in particular, the $\ln g$ piece does not depend on the R\'enyi index $n$~\cite{PasqualeCalabrese_2004,Calabrese_2009}. This provides an additional check of the boundary-entropy interpretation. Concretely, for an interval of length $\ell$ in a critical system, the R\'enyi entropy takes the form
\begin{equation}
S_n(\ell)=\frac{c}{6}\left(1+\frac{1}{n}\right)
\ln(\ell/\epsilon)
+2\ln g +s_n,
\label{eq:SM_Renyi_BCFT}
\end{equation}
where $\epsilon$ is a short-distance cutoff, $2\ln g$ are the boundary entropies associated with the two endpoints, and $s_n$ is an $n$-dependent nonuniversal constant.

As a starting point, for a Gaussian free-fermion state the integer R\'enyi entropy can be written in terms of correlation
eigenvalues $\{\nu_j\}$ as
\begin{equation}
S_n(A)=\frac{1}{1-n}\sum_j \ln\!\Big(\nu_j^n+(1-\nu_j)^n\Big),\qquad n\in\mathbb{Z}_{>0}.
\label{eq:SM_renyi_corr}
\end{equation}
In contrast to the von Neumann entropy, Eq.~\eqref{eq:SM_renyi_corr} contains no $\nu$-dependent prefactor multiplying the logarithm. As a result, a branch shift can only change $S_n$ by a purely imaginary \emph{constant} and cannot feed into the real-part scaling. Therefore, unlike the von Neumann case where enforcing conformal scaling effectively fixes the branch choice, at integer $n$ there is no analogous ``forced'' branch selection from conformal scaling; the following discussion therefore reports results based on a natural branch convention only.

Moreover, the polynomial combination $\nu^n+(1-\nu)^n$ probes the biorthogonal spectrum in a way that differs qualitatively from the von Neumann case. In particular, at even $n$ it is less sensitive to the sign/phase structure associated with correlation eigenvalues outside the Hermitian range $[0,1]$, and cancellations between the $\nu$ and $(1-\nu)$ contributions that are operative in the von Neumann entropy~\footnote{For instance, for sufficiently large $\nu_j>2$ the two terms in $S_j=-\nu_j\ln\nu_j-(1-\nu_j)\ln|1-\nu_j|$ can partially cancel.} need not be mirrored at even R\'enyi index. Empirically, we find that even-$n$ R\'enyi entropies do not recover the expected conformal scaling already at the trivial critical point. We therefore restrict the following analysis to odd-$n$ R\'enyi entropies.

With the above conventions in mind, we now evaluate Eq.~\eqref{eq:SM_renyi_corr} in the regime relevant to the topological critical line. For the bulk modes, where the correlation eigenvalues are real, one has $\nu_j^n+(1-\nu_j)^n>0$ for integer $n$ and arbitrary real $\nu_j$. The natural branch convention for the logarithm of a positive number then implies that the bulk contributions carry no imaginary part. The imaginary part, if present, must therefore originate solely from the edge-sector conjugate pair $\nu_\pm=\tfrac12\pm i\mathcal{I}$, in close analogy with the von Neumann case. It is convenient to define
\begin{equation}
A_n(\mathcal{I}):=\Big(\tfrac12+i\mathcal{I}\Big)^n+\Big(\tfrac12-i\mathcal{I}\Big)^n
=2\,\mathrm{Re}\Big(\tfrac12+i\mathcal{I}\Big)^n\in\mathbb{R},
\end{equation}
so that the pair contribution is
\begin{equation}
S_n^{\rm pair}(\mathcal{I})=\frac{2}{1-n}\ln A_n(\mathcal{I}).
\label{eq:SM_renyi_pair}
\end{equation}
At criticality, the edge parameter diverges as the gap closes, $\mathcal{I}\to\infty$. Writing
\begin{equation}
\tfrac12+i\mathcal{I} = r\,e^{i\theta},\qquad 
\theta=\arg(\tfrac12+i\mathcal{I})=\frac{\pi}{2}-\epsilon,\ \ \epsilon\to 0^+,
\end{equation}
we have
\begin{equation}
(\tfrac12+i\mathcal{I})^n=r^n e^{in\theta}
\quad\Rightarrow\quad
\arg\!\big[(\tfrac12+i\mathcal{I})^n\big]=\frac{n\pi}{2}-n\epsilon,
\qquad \epsilon\to 0^+ .
\end{equation}
Since $A_n(\mathcal{I})=2\,\mathrm{Re}(\tfrac12+i\mathcal{I})^n$, its sign is controlled by $\cos(n\theta)$.
For odd $n$, we have
\begin{equation}
A_n(\mathcal{I})=|A_n(\mathcal{I})|\,(-1)^{(n-1)/2}\qquad (n\ \text{odd},\ \mathcal{I}\to\infty).
\label{eq:SM_sign}
\end{equation}
Adopting the self-consistent branch choice $\ln(-1)=i\pi$, we obtain
\begin{equation}
\ln A_n(\mathcal{I})=\ln|A_n(\mathcal{I})|+i\pi\frac{n-1}{2}\qquad (n\ \text{odd},\ \mathcal{I}\to\infty).
\label{eq:SM_branch_An}
\end{equation}
Substituting Eq.~\eqref{eq:SM_branch_An} into Eq.~\eqref{eq:SM_renyi_pair} yields an $n$-independent imaginary entropy,
\begin{equation}
\mathrm{Im}\,S_n^{\rm pair}(\mathcal{I})=-\pi\qquad (n\ \text{odd},\ \mathcal{I}\to\infty),
\label{eq:SM_ImSn_pair}
\end{equation}
consistent with the interpretation of the imaginary constant as an endpoint (boundary-entropy) contribution.

It is also important to clarify the scope of this construction. The derivation above relies on the $\mathcal{I}\to\infty$ limit, which is naturally realized at the topological critical point where the gap closes. This behavior is not generic: for edge modes in a gapped phase, or for edge structures in higher-$\alpha$ models with longer-range hopping, we do not have $\mathcal{I}\to\infty$. In such cases the alternating-sign structure in Eq.~\eqref{eq:SM_sign} is absent, and we cannot identify a natural branch choice that yields an $n$-independent imaginary constant. 
This limitation is harmless from the BCFT viewpoint. The statement that the boundary-entropy contribution enters as an additive constant in the R\'enyi entropies is derived for \emph{local} critical points described by BCFT \cite{PasqualeCalabrese_2004,Calabrese_2009}. Away from criticality, or in the presence of additional long-range hopping that may invalidate the assumptions of the BCFT analysis, the R\'enyi-entropy structure need not follow the same BCFT pattern.

\section{Further evidence for $c=-2$ symmetry enriched criticality}
\label{sm6}

\subsection{Entanglement entropy scaling}

\begin{figure}[htbp]
	\centering
	\begin{minipage}[b]{0.32\linewidth}
		\panelH{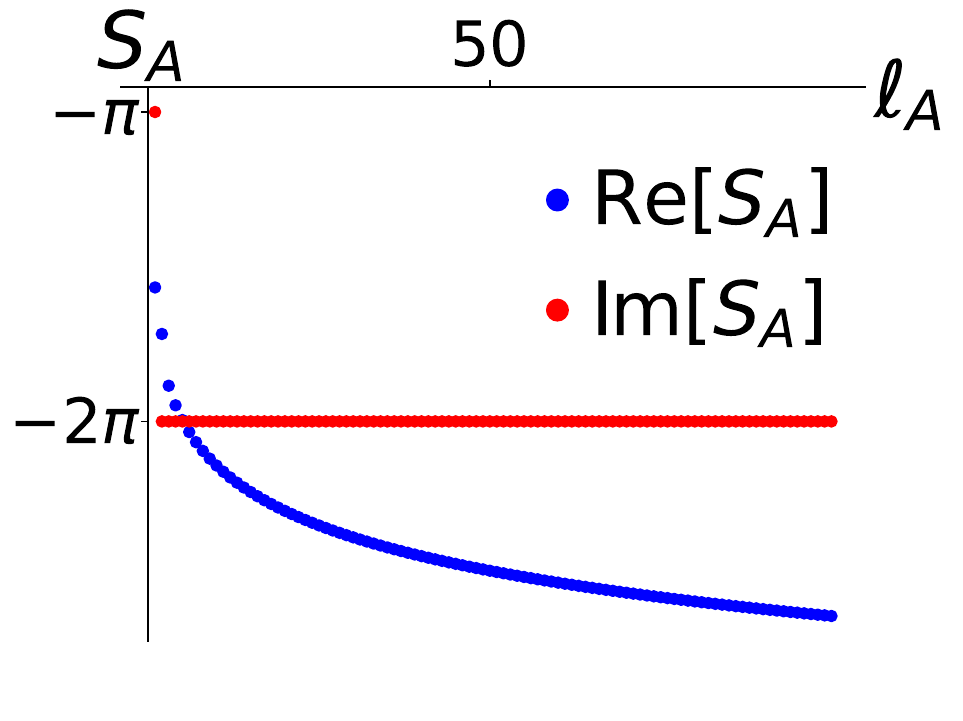}{4.3cm}{(a)}
	\end{minipage}
	\begin{minipage}[b]{0.32\linewidth}
		\panelH{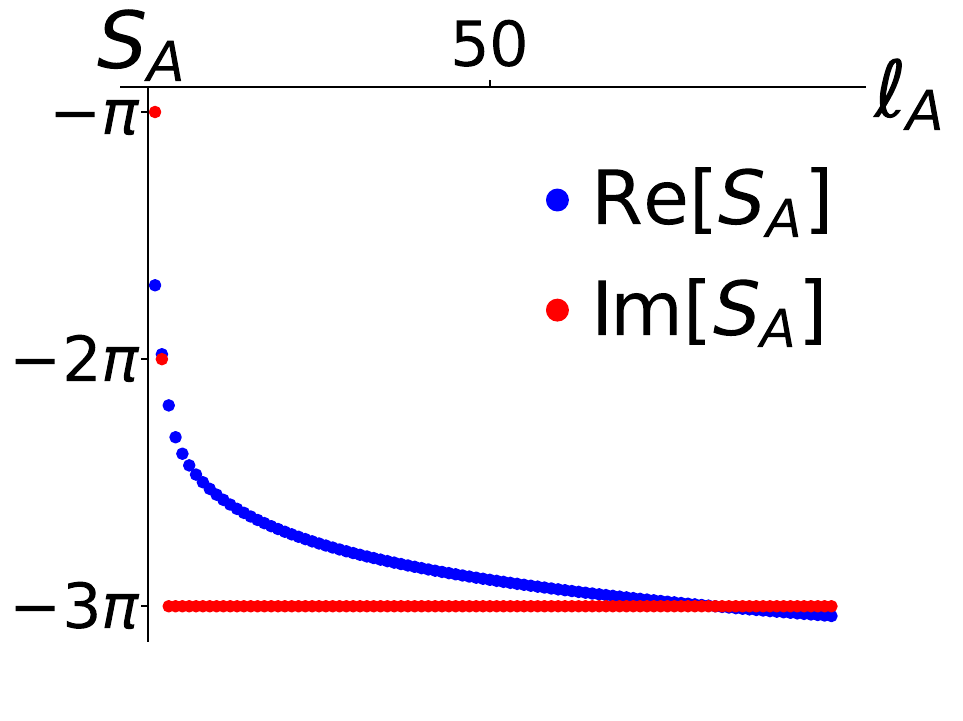}{4.3cm}{(b)}
	\end{minipage}
    \begin{minipage}[b]{0.32\linewidth}
    	\panelH{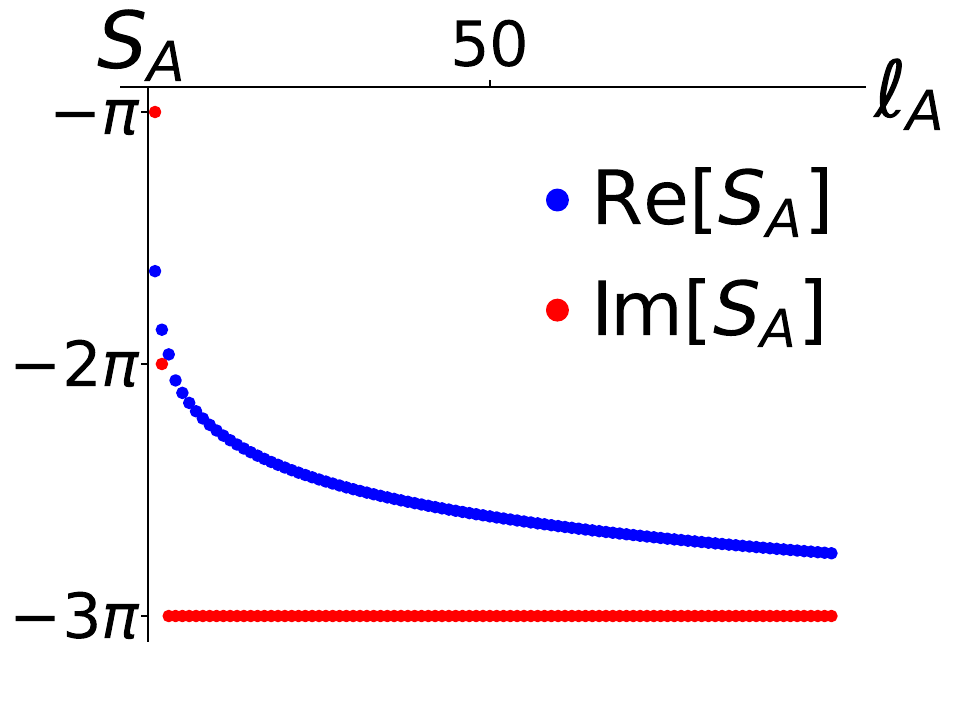}{4.3cm}{(c)}
    \end{minipage}\\
    \begin{minipage}[b]{0.32\linewidth}
    	\panelH{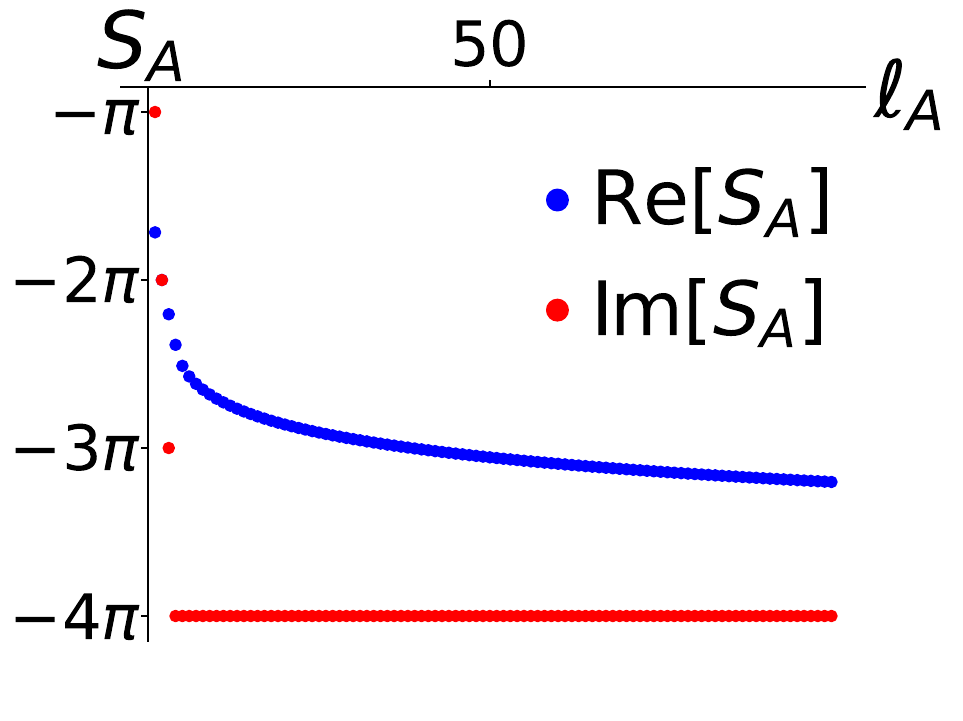}{4.3cm}{(d)}
    \end{minipage}
    \begin{minipage}[b]{0.32\linewidth}
    	\panelH{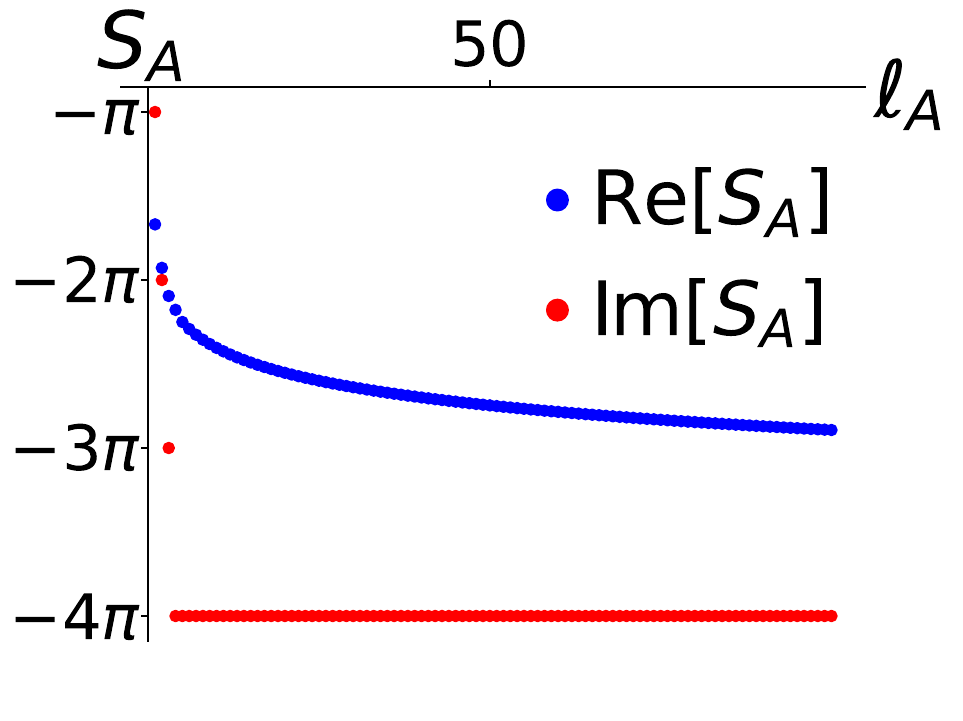}{4.3cm}{(e)}
    \end{minipage}
    \begin{minipage}[b]{0.32\linewidth}
		\panelH{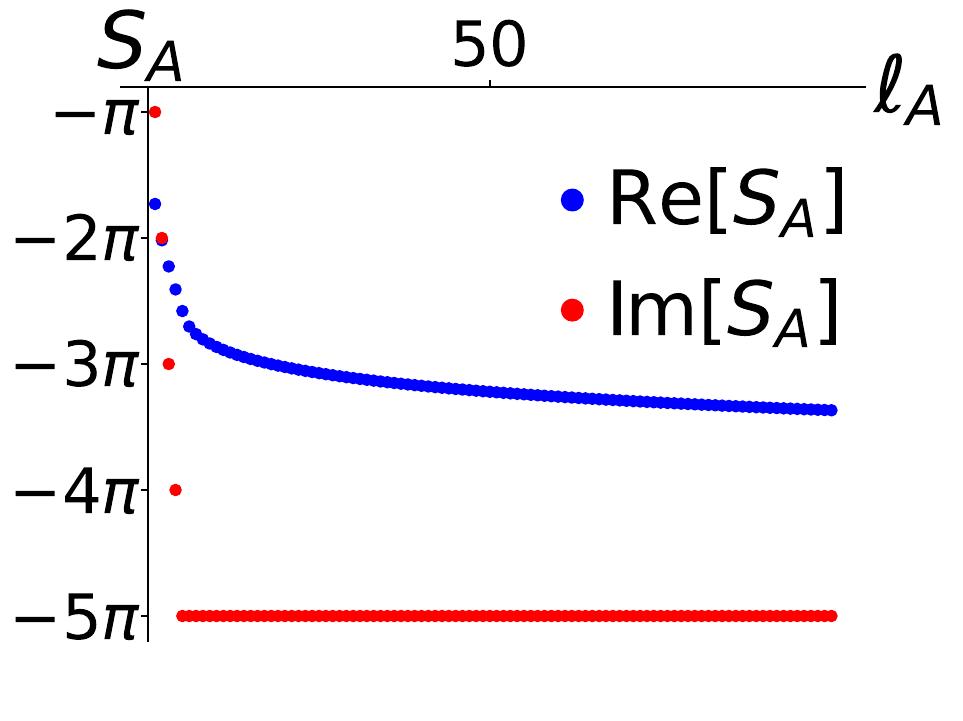}{4.3cm}{(f)}
	\end{minipage}
	\caption{PBC entanglement entropy at higher-$\alpha$, $L=10000$. Blue dots: $\operatorname{Re}[S_A]$; red dots: $\operatorname{Im}[S_A]$. Lower-winding QCPs use $(v,w,u)=(2,1,1-10^{-12})$; higher-winding QCPs use $(1,2,1-10^{-12})$. (a) $\alpha=3$ (lower): $c/3=-0.6644$. (b) $\alpha=3$ (higher): trim $\ell_A\le 10$ (SSE $\le 10^{-4}$); $c/3=-0.6668$. (c) $\alpha=4$ (lower): $c/3=-0.6658$. (d) $\alpha=4$ (higher): trim $\ell_A\le 14$ (SSE $\le 10^{-4}$); $c/3=-0.6677$. (e) $\alpha=5$ (lower): $c/3=-0.6683$. (f) $\alpha=5$ (higher): trim $\ell_A\le 17$ (SSE $\le 10^{-4}$); $c/3=-0.6690$. In all panels, $\operatorname{Im}[S_A]$ is step-like at small $\ell_A$ and saturates to $-\pi\Omega$ once all edge modes are covered, while $\operatorname{Re}[S_A]$ follows the Calabrese–Cardy scaling consistent with $c=-2$.}
	\label{fig:EE_pbc}
\end{figure}

We extend the periodic–boundary entanglement check of the central charge to higher-$\alpha$ cases not covered in the main text. Throughout this subsection we set $L=10^4$. For lower-winding critical points we use $(v,w,u)=(2,1,1-10^{-12})$, while for higher-winding critical points we use $(1,2,1-10^{-12})$. The real part of the entropy is fitted to the Calabrese-Cardy form
\begin{equation}
	\operatorname{Re}[S_A^{\rm PBC}](\ell_A)=\frac{c}{3}\ln \left[ \sin \left(\frac{\pi \ell_A}{L}\right)\right]+s_0
\end{equation}
The imaginary part under PBC is step-like in $\ell_A$: it drops by $-\pi$ each time an edge mode is included and then saturates once all $\Omega$ modes are covered. Eventually, $\operatorname{Im} S_A^{\rm PBC}=-\pi\Omega$ for a QCP with winding number $\Omega$.

In the small-$\ell_A$ regime we observe a pronounced UV deviation. For lower-winding QCPs, this originates from unit cell granularity: a single unit cell captures at most one edge mode, so one must have $\ell_A>\Omega$ to include all $\Omega$ modes; therefore it is not surprising to see the UV deviation before this threshold. Furthermore, at the higher-winding QCPs, we have an additional edge mode generated by generalized mass inversion, thus with a finite decay length $\xi$, so the overlap with the cut decays smoothly over a scale $\sim\xi$; correspondingly, the deviation extends over a longer range. To obtain a clean Calabrese–Cardy fit, we exclude a minimal subset of contaminated UV points: for lower-winding QCPs we remove the first $N_{\rm edge}$ data points, while for higher-winding QCPs we choose the smallest trim such that the sum of squared residuals satisfies $\mathrm{SSE}\le 10^{-4}$.

As shown in Fig.~\ref{fig:EE_pbc}, across all higher-$\alpha$ cases the real part of the PBC entanglement follows the Calabrese–Cardy form with slope $c/3\simeq -2/3$ once a minimal UV trim is applied. The imaginary part exhibits the expected step-to-saturation behavior and ultimately equals $-\pi\Omega$. Taken together, these checks confirm $c=-2$ for the higher-$\alpha$ critical points considered here.

One might also ask for an entanglement-based CFT check under OBC. 
We do not include an OBC entanglement check here for the following reason. 
For an open chain, the standard contiguous subsystem \(A=[1,\ell_A]\) is not invariant under inversion and therefore breaks the \(PT\) symmetry (\(P(A)\neq A\)). 
In contrast, under PBC translational invariance allows the inversion center to be chosen freely; one may therefore choose \(A\) to be inversion-symmetric by placing the inversion center at the midpoint of the subsystem, so that \(P(A)=A\).
Consequently, under OBC the restricted correlator \(C_A\) does not necessarily satisfy the \(PT\) constraint discussed in Sec.~\ref{subsec:symmetry-quartet}, and its spectrum is not guaranteed to exhibit the \(PT\)-enforced complex-conjugation pairing. 
Without this symmetry structure, a self-consistent branch choice for the entanglement entropy becomes ambiguous, and we thus restrict the entanglement analysis to PBC.

\subsection{Ground-state energy scaling}

\begin{figure}[htbp]
	\centering
	\begin{minipage}[b]{0.24\linewidth}
		\panelH{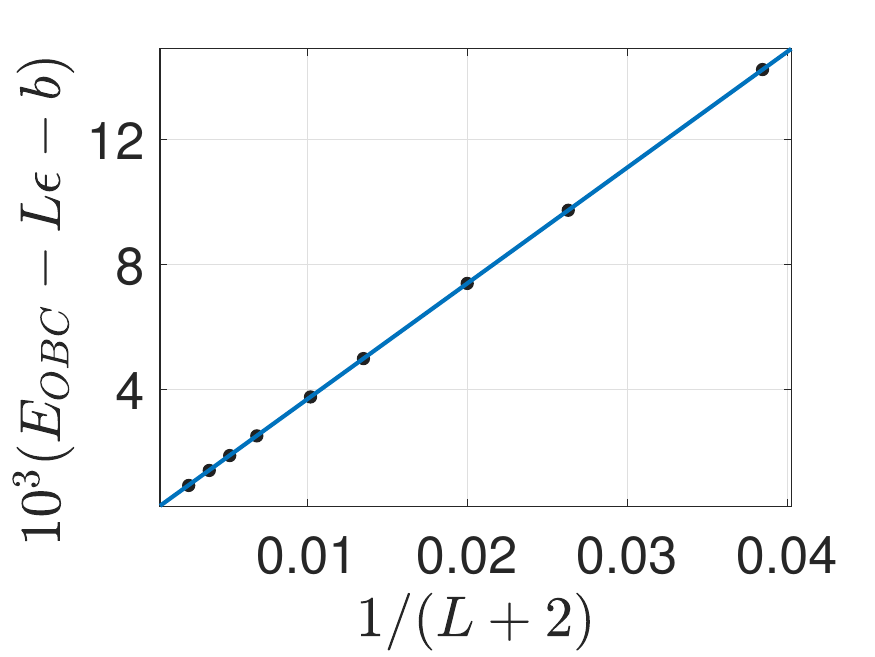}{3.2cm}{(a)}
	\end{minipage}
	\begin{minipage}[b]{0.24\linewidth}
		\panelH{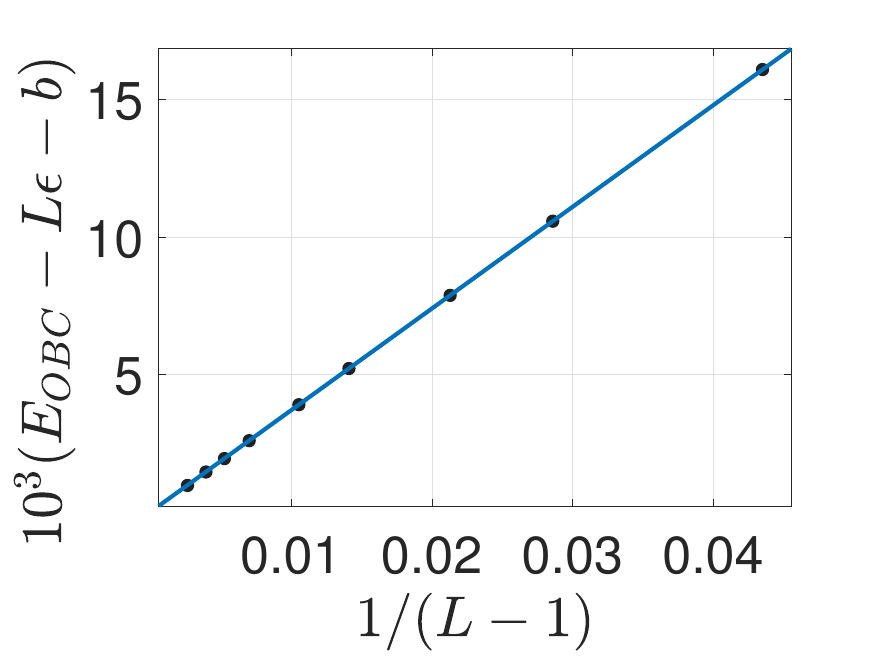}{3.2cm}{(b)}
	\end{minipage}
	\begin{minipage}[b]{0.24\linewidth}
		\panelH{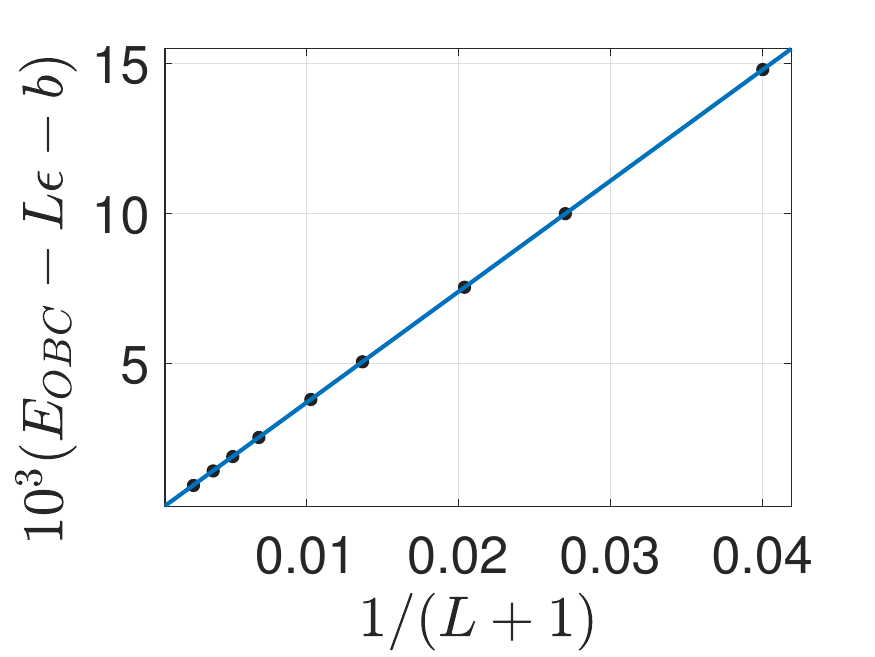}{3.2cm}{(c)}
	\end{minipage}
	\begin{minipage}[b]{0.24\linewidth}
		\panelH{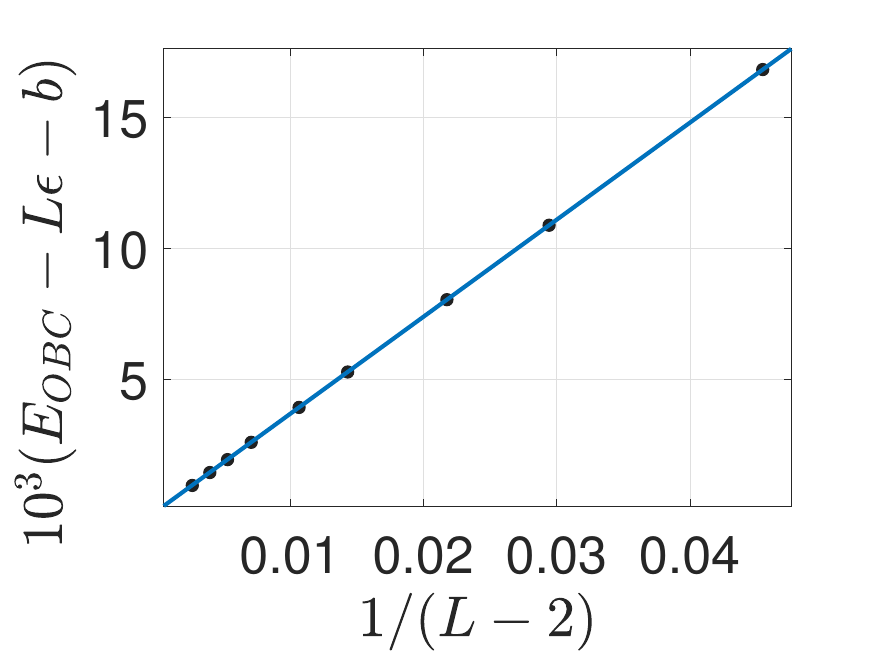}{3.2cm}{(d)}
	\end{minipage}
	\caption{OBC Casimir scaling.
		Numerical fit slopes \(A\) of \(E_0^{\rm OBC}\) versus \(1/L_{\rm eff}\):
		(a) $\alpha=1$ trivial QCP: $A=0.3701466$;
		(b) $\alpha=1$ topological QCP: $A=0.3702412$;
		(c) $\alpha=2$ lower winding QCP: $A=0.3701497$;
		(d) $\alpha=2$ higher winding QCP: $A=0.3702569$.
		All slopes are consistent with the $c=-2$ expectation for $v_F=\sqrt{2}$.}
	\label{fig:casimir}
\end{figure}

We begin from the standard Casimir forms for open and periodic chains,
\begin{equation}
E_0^{\rm OBC}(L)=L\,\epsilon+b-\frac{\pi v_F c}{24\,L}+\mathcal{O}(L^{-2}),\qquad
E_0^{\rm PBC}(L)=L\,\epsilon-\frac{\pi v_F c}{6\,L}+\mathcal{O}(L^{-2}),
\end{equation}
where $\epsilon$ is the bulk energy density, $b$ the order-$L^0$ boundary constant, and $v_F$ the bulk velocity (fixed to $v_F=\sqrt{2}$).
Following boundary CFT and surface critical analyses, leading boundary corrections can be absorbed into an \emph{extrapolation length}—a microscopic shift of the effective boundary position—implemented by replacing $L\to L_{\rm eff}=L+\Delta_L$ in the Casimir term \cite{Diehl:1996kd,Stéphan_2013}. This yields the working-fit form,
\begin{equation}
\label{eq:casimir}
E_0^{\rm OBC}(L)=L\,\epsilon+b-\frac{\pi v_F c}{24\,(L+\Delta_L)}\,,
\end{equation}
with a non-universal $\Delta_L=\mathcal O(1)$ that depends on the boundary class.

Using Eq.~\eqref{eq:casimir}, the expected $1/L$ slopes follow immediately. For $c=-2$ and $v_F=\sqrt{2}$,
\begin{equation}
A_{\rm OBC}
=\frac{\pi v_F(-c)}{24}
\approx 0.370240,\qquad
A_{\rm PBC}=\frac{\pi v_F(-c)}{6}\approx 1.480961 .
\end{equation}
For reference, the PBC fit reported in the main text [Fig.~2(d)] yields $A_{\rm PBC}=1.4807$, which is in excellent agreement with $c=-2$. A single non-universal, $\mathcal O(1)$ extrapolation length per boundary class brings the OBC data onto this line:
\begin{equation}
\alpha=1:\ \Delta_L^{\rm triv}=+2,\ \ \Delta_L^{\rm topo}=-1;\qquad
\alpha=2:\ \Delta_L^{\rm lower}=+1,\ \ \Delta_L^{\rm higher}=-2 .
\end{equation}
With these shifts, the fits in Fig.~\ref{fig:casimir} align with the $c=-2$ prediction.

\section{Edge-mode solutions at non-Hermitian criticality}
\label{sm7}
\subsection{Lattice derivation}

In this subsection we derive the edge modes directly from the lattice eigenequations. For the critical nH SSH chain under OBC, the left-boundary and bulk equations can be written as
\begin{align}
    v\psi_{B,0} &= (E-iu)\psi_{A,0}, \label{eq:SM_edge_eq1}\\
    v\psi_{A,n} + w\psi_{A,n+1} &= (E+iu) \psi_{B,n}, \qquad (n\ge0), \label{eq:SM_edge_eq2}\\
    w\psi_{B,n} + v\psi_{B,n+1} &= (E-iu) \psi_{A,n+1}. \qquad (n\ge0) \label{eq:SM_edge_eq3}
\end{align}
Substituting the exponential ansatz $(\psi_{A,n},\psi_{B,n})=\beta^n(\psi_{A,0},\psi_{B,0})$ into \eqref{eq:SM_edge_eq2}--\eqref{eq:SM_edge_eq3} gives
\begin{align}
    (v+w\beta)\psi_{A,0} &= (E+iu)\psi_{B,0}, \label{eq:SM_edge_ans1}\\
    (v+w\beta^{-1})\psi_{B,0} &= (E-iu)\psi_{A,0}. \label{eq:SM_edge_ans2}
\end{align}
Comparing \eqref{eq:SM_edge_ans2} with the boundary condition \eqref{eq:SM_edge_eq1}, one finds that a nontrivial decaying solution forces $\psi_{B,0}=0$. This immediately gives $E=iu$, and \eqref{eq:SM_edge_ans1} further implies
\begin{equation}
    v+w\beta=0 \quad\Rightarrow\quad \beta=-\frac{v}{w}.
\end{equation}
Normalizability $|\beta|<1$ gives $|v|<|w|$, consistent with the field-theory generalized mass inversion criterion. 
Notice that for the nH SSH model the critical condition is $|w|-|v|=u>0$, so one can still have $|\beta|=|v/w|<1$ at criticality. 
The right-end solution follows similarly, with $\psi_A=0$ and $E=-iu$.

For the $\alpha$-extended nH SSH model, the left-boundary truncation of the longer-range hopping leads to the boundary equations
\begin{align}
    0 &= (E-iu)\psi_{A,0}, \label{eq:SM_aedge_eq0}\\
    v\psi_{B,0} &= (E-iu)\psi_{A,1}, \label{eq:SM_aedge_eq1}
\end{align}
together with the bulk recursion
\begin{align}
    v\psi_{A,n+1} + w\psi_{A,n+2} &= (E+iu) \psi_{B,n}, \qquad (n\ge0), \label{eq:SM_aedge_eq2}\\
    w\psi_{B,n} + v\psi_{B,n+1} &= (E-iu) \psi_{A,n+2}. \qquad (n\ge0) \label{eq:SM_aedge_eq3}
\end{align}
Equation \eqref{eq:SM_aedge_eq0} shows that $\psi_{A,0}$ is completely decoupled from the bulk and hence produces an exact boundary eigenstate with $E=iu$ (e.g. $\psi_{A,0}=1$ and all other amplitudes zero). To obtain the extended edge mode, we again substitute an exponential form for the active sites, $(\psi_{A,n+1},\psi_{B,n})=\beta^{n}(\psi_{A,1},\psi_{B,0})$, which gives
\begin{align}
    (v+w\beta)\psi_{A,1} &= (E+iu)\psi_{B,0}, \label{eq:SM_aedge_ans1}\\
    (v+w\beta^{-1})\psi_{B,0} &= (E-iu)\psi_{A,1}. \label{eq:SM_aedge_ans2}
\end{align}
Comparing \eqref{eq:SM_aedge_ans2} with the boundary condition \eqref{eq:SM_aedge_eq1} again enforces $\psi_{B,0}=0$ for a nontrivial decaying solution, which yields $E=iu$ and
\begin{equation}
    v+w\beta=0 \quad\Rightarrow\quad \beta=-\frac{v}{w},
\end{equation}
with normalizability requiring $|v|<|w|$. Altogether, the $\alpha$-extended model supports $(\alpha-1)$ decoupled edge-mode pairs from the truncated boundary sites, in addition to one edge-mode pair associated with the generalized mass inversion.

\subsection{Higher-$\alpha$ continuum analysis}

We start from the block Hamiltonian of the $\alpha$-non-Hermitian SSH model
\begin{equation}
\mathcal{H}(k)=
\begin{bmatrix}
iu & v_k \\
v_k^* & -iu
\end{bmatrix},
\qquad
v_k = v\,e^{-i(\alpha-1)k} - w\,e^{-i\alpha k},
\end{equation}
and pass to a continuum description via the non-local map  
\begin{equation}
e^{ik}\ \Rightarrow\ (\partial_x+1).
\end{equation}
We use this replacement---rather than expanding $v_k$ to $\alpha$-th order in $k$---because finite Taylor series mix different hopping ranges into the same derivative coefficients, letting near-neighbor terms renormalize those that should represent longer-range physics. In contrast, the map $e^{-ink}\mapsto(\partial_x+1)^n$ keeps each range separate and ensures that this contamination does not occur. The field-theory Hamiltonian then takes the form
\begin{equation}
\mathcal{H}(x)=
\begin{bmatrix}
iu & D_x^\dagger \\
D_x & -iu
\end{bmatrix},
\qquad
D_x = v(\partial_x+1)^{\alpha-1} - w(\partial_x+1)^\alpha .
\end{equation}
As on the lattice, the eigenvalue equations decouple at $E=\pm iu$. For a left-edge mode we set $E=iu$ and $\psi_B\equiv 0$, which gives the scalar boundary equation,
\begin{equation}
(\partial_x+1)^{\alpha-1}\,(v-w - w\,\partial_x)\,\psi_A(x)=0.
\label{eq:ft-edge}
\end{equation}
We now substitute the exponential ansatz $\psi_A(x)\sim e^{\beta x}$ into Eq.~\eqref{eq:ft-edge}, 
noting that only solutions with $\beta<0$ are normalizable.
\begin{equation}
(\beta+1)^{\alpha-1}\,\bigl[(v-w)-w\,\beta\bigr]=0,
\end{equation}
whose roots are
\begin{equation}
\beta=-1 \quad (\text{$\alpha-1$-fold}), \qquad \beta=\frac{v-w}{w}.
\end{equation}
It follows immediately that the number of edge mode pairs is $\alpha-1$ for $v>w$ and $\alpha$ for $w>v$, in agreement with the lattice result.

The $(\alpha-1)$-fold solution connects directly to the lattice picture. In the $\alpha$-nH SSH chain, the minimal hopping leg spans $\alpha$ sites; with an open boundary this leaves $\alpha-1$ sites that decouple from the bulk and thus form edge modes. 
In the continuum description, the lattice translation factor $e^{-i(\alpha-1)k}$ maps to $(\partial_x+1)^{\alpha-1}$, yielding edge solutions that mirror the same physics. After removing this factor, the residual equation
\begin{equation}
\bigl[(v-w)-w\,\partial_x\bigr]\psi_A=0
\end{equation}
is precisely the generalized mass-inversion condition discussed in the main text.

To examine whether a kinetic-inversion–type picture applies, we consider the $\alpha=2$ case. The edge equation is
\begin{equation}
\bigl[-w\,\partial_x^2+(v-2w)\,\partial_x+(v-w)\bigr]\psi_A=0.
\end{equation}
On the line $v=w$, this reduces to the kinetic-inversion case discussed previously~\cite{verresen2020topologyedgestatessurvive}. 
Away from that line, however, the existence of an edge mode is not captured by a single sign flip (mass or kinetic inversion): for example, when $v>2w$ a negative root $\beta=-1$ still persists even though neither the effective mass nor the kinetic coefficient changes sign at the boundary. 
Because the second-order ordinary differential equation involves multiple independent coefficients, a one-parameter inversion picture is too restrictive; the appropriate criterion is the full characteristic equation.

\subsection{Interface geometry}

Recall the field-theory equations in the nH bulk
\begin{align}
(-\partial_x + m_2)\psi_B(x) &= (E - i u)\psi_A(x), \nonumber\\
(\partial_x + m_2)\psi_A(x)  &= (E + i u)\psi_B(x),
\end{align}
and in the Hermitian bulk
\begin{align}
(-\partial_x + m_1)\psi_B(x) &= E\psi_A(x), \nonumber\\
(\partial_x + m_1)\psi_A(x)  &= E\psi_B(x).
\end{align}
We work at the topological QCP on the non-Hermitian side, $u=m_2$. Assume exponential behavior $\psi_{A,B}\sim e^{\kappa_2 x}$ for $x>0$ (non-Hermitian side) and $\psi_{A,B}\sim e^{-\kappa_1 x}$ for $x<0$ (Hermitian side); an interface-pinned mode then requires $\kappa_{1,2}<0$. This leads to the sublattice ratios
\begin{align}
\frac{\psi_A}{\psi_B}
&= \frac{E + i u}{\kappa_2 + m_2}
= \frac{-\kappa_2 + m_2}{E - i u}, \nonumber\\
\frac{\psi_A}{\psi_B}
&= \frac{E}{-\kappa_1 + m_1}
= \frac{\kappa_1 + m_1}{E}. \label{eq:ratio}
\end{align}
On both sides the bulk dispersion relations are
\begin{align}
m_1^2 - \kappa_1^2 &= E^2, \nonumber\\
m_2^2 - \kappa_2^2 &= E^2 + u^2.
\end{align}
With $u=m_2$ this gives
\begin{align}
\kappa_2 &= iE \quad (\text{choosing the branch consistent with decay}), \nonumber\\
\kappa_1 &= -\sqrt{m_1^2 - E^2}. \label{eq:kappa}
\end{align}
Continuity of $\psi_A/\psi_B$ at the interface requires
\begin{equation}
E(-\kappa_2 + m_2) = (E - i u)(\kappa_1 + m_1).
\end{equation}
Substituting \eqref{eq:kappa} yields
\begin{align}
E(-iE + m_2) &= (E - iu)\bigl(-\sqrt{m_1^2 - E^2} + m_1\bigr),\nonumber\\
E(-iE + m_2) - (E - iu)m_1 &= -(E - iu)\sqrt{m_1^2 - E^2}
\end{align}
Squaring both sides and simplifying gives
\begin{equation}
-i4E^3 u + i2E(E^2 + u^2)m_1 = 0,
\end{equation}
which leads to
\begin{equation}
E = \sqrt{\frac{u^{2} m_1}{2u - m_1}}
= i u\sqrt{\frac{-m_1}{2u - m_1}}. \label{eq:solution}
\end{equation}
From the solution we can write $E=i a u$.  
For $m_1<0$ one finds $a\in[0,1]$, giving an interface edge mode (mass inversion).  
As $m_1\to 0$, the mode merges into the bulk on both sides.

By contrast, when $m_1>2u$ we have
\begin{equation}
a = \sqrt{\frac{m_1}{m_1-2u}} > 1.
\end{equation}
Substituting $E=i a u$ and the $\kappa$ solutions \eqref{eq:kappa} into the sublattice ratios \eqref{eq:ratio} gives
\begin{align}
\left(\frac{\psi_A}{\psi_B}\right)_{R}
&= \frac{i a u + i u}{-a u + u}
= i\frac{a+1}{1-a}
= -i\frac{a+1}{a-1} \quad (<0\ \text{imag}),\nonumber\\
\left(\frac{\psi_A}{\psi_B}\right)_{L}
&= \frac{i a u}{\sqrt{m_1^2 + a^2 u^2}+ m_1}
= i\frac{a u}{\sqrt{m_1^2 + a^2 u^2}+ m_1} \quad (>0\ \text{imag}).
\end{align}
Thus the right ratio is purely negative imaginary while the left ratio is purely positive imaginary; they cannot match, so the boundary condition fails. This shows that the $m_1>2u$ branch (with $a>1$) is an extraneous root generated by squaring the equation. Interface edge modes occur only when mass inverse at the interface.

\begin{figure}[htbp]
	\centering
	\begin{minipage}[b]{0.32\linewidth}
		\panelH{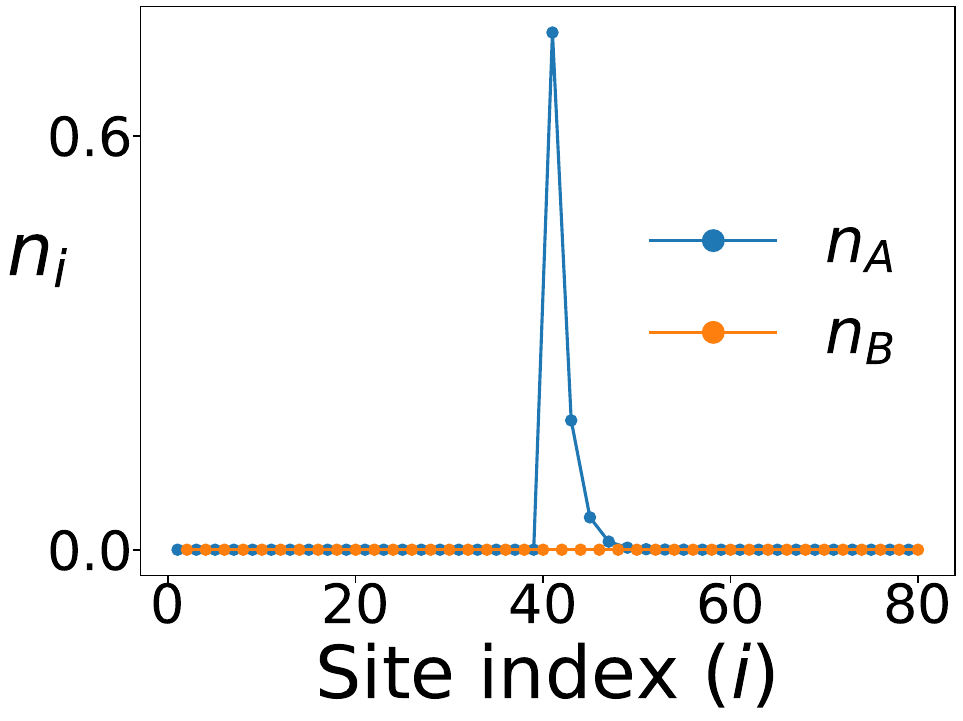}{4.3cm}{(a)}
	\end{minipage}
	\begin{minipage}[b]{0.32\linewidth}
		\panelH{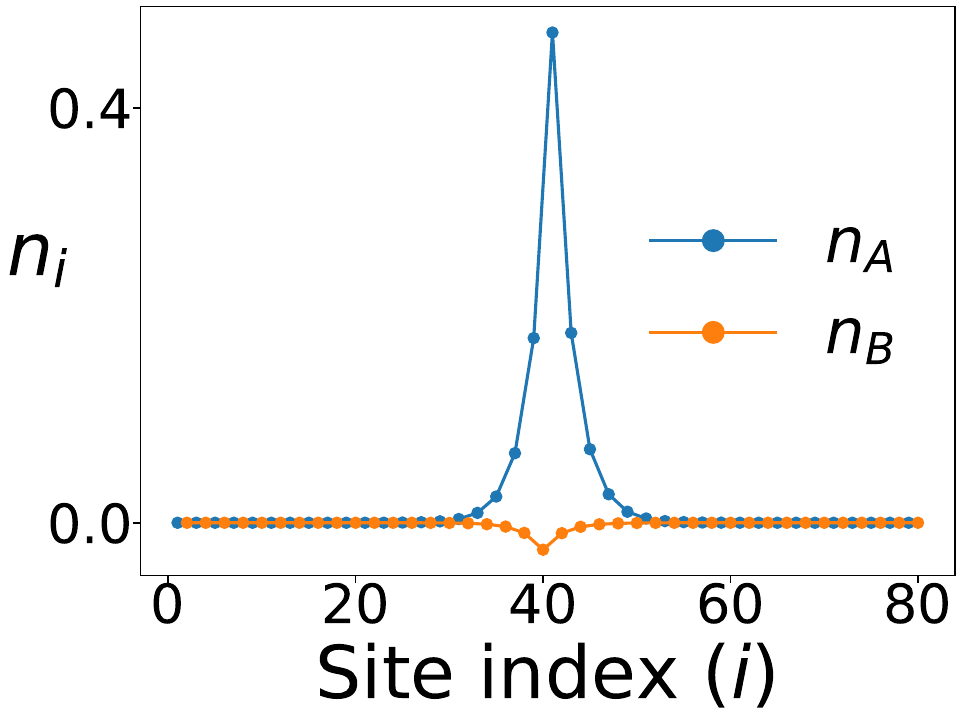}{4.3cm}{(b)}
	\end{minipage}
	\begin{minipage}[b]{0.32\linewidth}
		\panelH{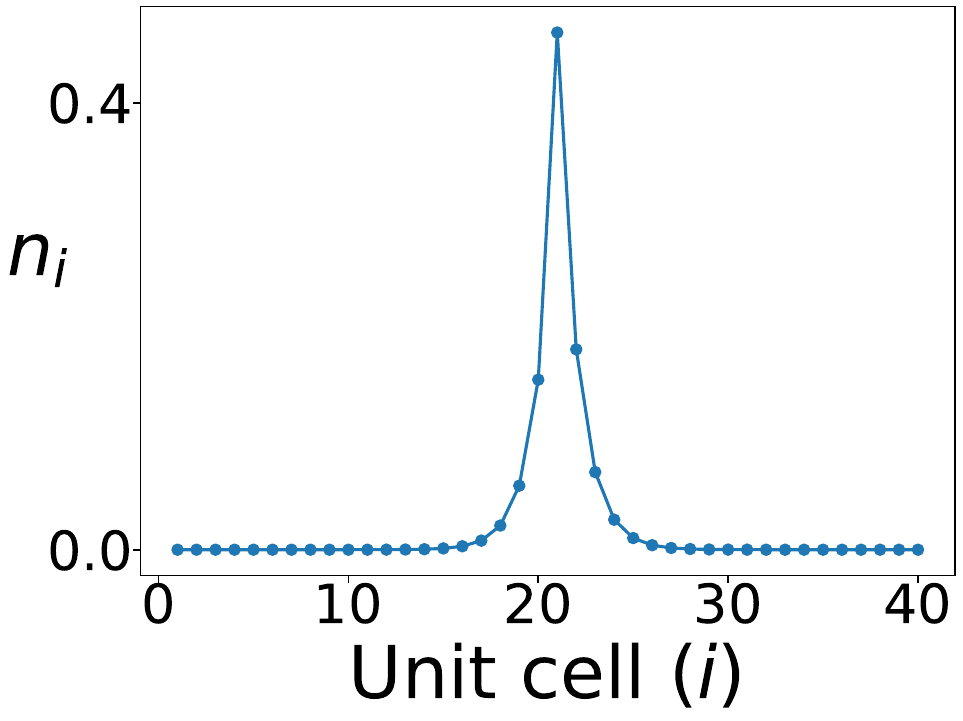}{4.3cm}{(c)}
	\end{minipage}\\
	\begin{minipage}[b]{0.32\linewidth}
		\panelH{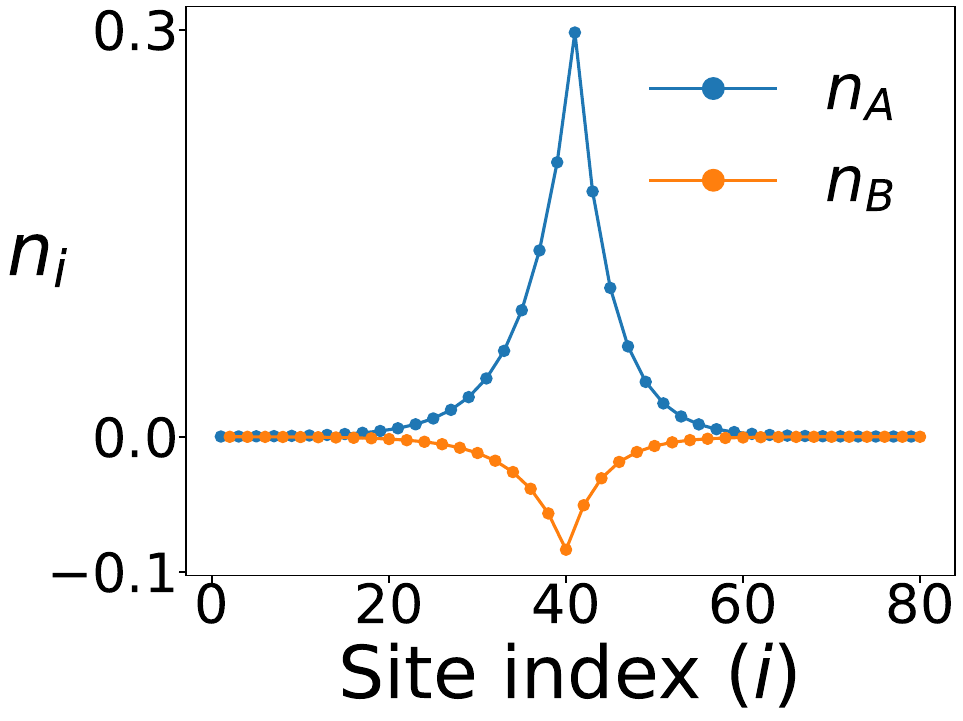}{4.3cm}{(d)}
	\end{minipage}
	\begin{minipage}[b]{0.64\linewidth}
		\panelH{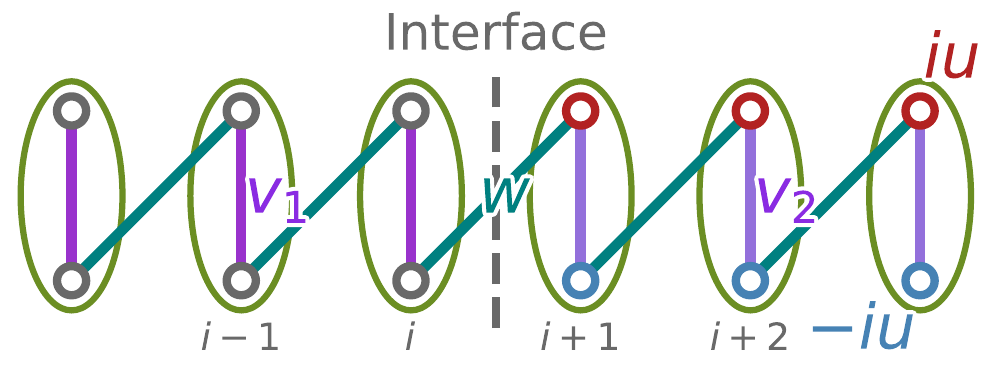}{4.3cm}{(e)}
	\end{minipage}
	\caption{Edge mode pinned at a Hermitian–non-Hermitian interface. The nH side is fixed at $(v_2,w,u)=(0.5,1,0.5)$.
		Unit cells $i=1\text{–}20$ (sites $1\text{–}40$) are Hermitian; $i=21\text{–}40$ (sites $41\text{–}80$) are nH.
		Panels (a,b,d) show particle densities with $n_i = \bra{L_{edge}}c_i^\dagger c_i\ket{R_{edge}}$; panel (c) shows the unit cell density $n_i=n_{i,A}+n_{i,B}$.
		(a) "Infinite-mass" left side: Hermitian $(v_1,w)=(100,1)$. The edge mode reduces to the usual OBC profile.
		(b) Mass inversion: Hermitian $(v_1,w)=(1.5,1)$. An edge mode is pinned at the interface; the $B$-sublattice density is negative.
		(c) Unit cell representation for (b): recovers the familiar edge mode and reproduces Fig.~4(b).
		(d) Near gap closing on the Hermitian side: Hermitian $(v_1,w)=(1.1,1)$. The interface mode hybridizes with both continua and, as the Hermitian gap closes, fully merges into the bulk and delocalizes.
		(e) Lattice schematic of the interface: Hermitian couplings $v_1,w$ on the left; nH couplings $v_2,w$ with on-site $\pm i u$ on the right. The dashed line marks the interface.
	}
	\label{fig:interface}
\end{figure}

These analytic results are borne out by the numerics. In the limit $m_1\to -\infty$, Eq.~\eqref{eq:solution} yields $E=iu$. From \eqref{eq:kappa} and \eqref{eq:ratio}, the interface mode has decay rates $\kappa_2=-u$ and $\kappa_1\to -\infty$, with a vanishing $B$-sublattice amplitude. The solution therefore reduces to the OBC edge pinned on the non-Hermitian side, in agreement with the numerical profile shown in Fig.~\ref{fig:interface}(a). Physically, this is natural—the infinite-mass boundary is equivalent to vacuum.

For finite negative $m_1$, the edge mode remains pinned at the interface but now with a finite decay length on both sides, which is visible in Fig.~\ref{fig:interface}(b). An additional feature appears here: the plotted $B$-sublattice amplitude is negative. This can be explained by the imaginary sublattice ratio $\psi_A/\psi_B\sim+i$ as discussed above. The amplitude is computed in the biorthogonal basis ($\psi_L^*\psi_R$, effectively $\psi^2$ here), thus we obtain a negative $B$-sublattice amplitude.

The sign structure is physically consistent: the on-site imaginary potentials have opposite signs on $A/B$, so for a positive imaginary energy the negative $B$-sublattice amplitude aligns with the negative imaginary $B$-sublattice potential in the biorthogonal expectation, yielding a net positive imaginary contribution. If the configuration is reversed (non-Hermitian part on the left, Hermitian on the right), the roles swap and the energy becomes negative imaginary, producing the opposite sign pattern. We can also recover the familiar representation by computing the unit cell amplitude (Fig.~\ref{fig:interface}(c)), which reproduces the plot shown in the main text Fig.~4(b).

Finally, as $m_1\to 0^{-}$ the analytic solution enforces $\kappa_{1,2}\to 0^{-}$ and $E\to 0$, so the edge state merges into the bulk on both sides. The numerical profiles in Fig.~\ref{fig:interface}(d) confirm this: the decay lengths diverge on both sides. This contrasts with a purely Hermitian interface, where the decay length is set solely by local parameters; here, via $E$ and \eqref{eq:kappa}, the two bulks are coupled, so localization depends on both sides of the interface.

In summary, although the simple decoupling argument no longer applies at an interface, the generalized mass-inversion picture remains valid and continues to govern the existence of pinned edge states.

\emph{Lattice calculation}---For completeness, the same problem can be formulated directly in the lattice model. We set up the lattice interface as in Fig.~\ref{fig:interface}(e): the left side is the Hermitian SSH chain, and the right side is the nH SSH chain at its topological QCP. We label the unit cell adjacent to the interface by $i$. The single-particle equations near the interface are
\begin{align}
-w\psi_{i-1}^B + v_1\psi_i^B &= E\psi_{i}^A,\nonumber\\
v_1\psi_i^A - w\psi_{i+1}^A &= E\psi_i^B,\nonumber\\
-w\psi_i^B + v_2\psi_{i+1}^B &= (E-iu)\psi_{i+1}^A,\nonumber\\
v_2\psi_{i+1}^A - w\psi_{i+2}^A &= (E+iu)\psi_{i+1}^B, \label{eq:inter}
\end{align}
In each bulk we adopt an exponential ansatz with a fixed sublattice ratio:
\begin{align}
\frac{\psi^{B}_n}{\psi^{A}_n} &= \left(\frac{\psi^{B}}{\psi^{A}}\right)_{L} = \text{const.}, 
& \qquad \psi^{A,B}_{n-1} &= \beta_{L}\psi^{A,B}_{n}, 
& \qquad n &\in \text{left},\nonumber\\
\frac{\psi^{B}_n}{\psi^{A}_n} &= \left(\frac{\psi^{B}}{\psi^{A}}\right)_{R} = \text{const.}, 
& \qquad \psi^{A,B}_{n+1} &= \beta_{R}\psi^{A,B}_{n}, 
& \qquad n &\in \text{right}.
\end{align}
For an edge mode pinned at the interface we require $|\beta_{1,2}|<1$. Substitute the ansatz into the bulk eigen–equations to obtain the recursion relations
\begin{align}
\left(\frac{\psi^B}{\psi^A}\right)_{L}
&= \frac{E}{v_1 - w\beta_L}
= \frac{v_1 - w\beta_L^{-1}}{E},\nonumber\\
\left(\frac{\psi^B}{\psi^A}\right)_{R}
&= \frac{E - i u}{v_2 - w\beta_R^{-1}}
= \frac{v_2 - w\beta_R}{E + i u}.
\end{align}
Equating the two forms on each side, these can be rearranged into the bulk dispersions
\begin{align}
E^{2} &= v_1^{2} + w^{2} - v_1 w\left(\beta_L + \beta_L^{-1}\right),\nonumber\\
E^{2} + u^{2} &= v_2^{2} + w^{2} - v_2 w\left(\beta_R + \beta_R^{-1}\right).\label{eq:disp}
\end{align}
These relations are the lattice counterparts of the continuum bulk dispersions \eqref{eq:kappa}. For the interface boundary condition, we start from the eigen-equations \eqref{eq:inter} across the interface and substitute the fixed sublattice ratios to obtain
\begin{align}
-w\psi_{i+1}^A
&= \Big[E - v_1\left(\tfrac{\psi^A}{\psi^B}\right)_{L}\Big]\psi_i^B, \nonumber\\
-w\psi_i^B
&= \Big[E - i u - v_2\left(\tfrac{\psi^B}{\psi^A}\right)_{R}\Big]\psi_{i+1}^A.
\end{align}
Eliminating the ratio gives the boundary-matching condition:
\begin{equation}
\frac{E - v_1\left(\tfrac{\psi^A}{\psi^B}\right)_{L}}{-w}
=
\frac{-w}{E - i u - v_2\left(\tfrac{\psi^B}{\psi^A}\right)_{R}}.\label{eq:bc}
\end{equation}
In the long-wavelength limit, the sublattice ratio varies negligibly across a unit cell, so the lattice condition \eqref{eq:bc} reduces to the continuum boundary condition used in the field-theory treatment.

Upon here, we have all the equations needed to solve the interface problem: together, Eqs.~\eqref{eq:disp} and \eqref{eq:bc} provide three independent equations for the three unknowns $E, \beta_L, \beta_R$. However, while the ansatz is simple, solving explicitly for $\beta_{L,R}$ and $E$ leads to cumbersome algebra with no clean closed form. We therefore present the lattice formulation solely as a reference and numerical validation of the continuum analysis.


\end{appendix}



\bibliography{main}
\nolinenumbers

\end{document}